\documentclass[fleqn,usenatbib]{mnras}

\usepackage[T1]{fontenc}

\DeclareRobustCommand{\VAN}[3]{#2}
\let\VANthebibliography\thebibliography
\def\thebibliography{\DeclareRobustCommand{\VAN}[3]{##3}\VANthebibliography}

\usepackage{graphicx}	
\usepackage{amsmath}	
\usepackage{amssymb}	
\usepackage{multirow,bigdelim}
\usepackage{cuted}
\usepackage{hyperref}
\usepackage{url}
\usepackage[dvipsnames]{xcolor}
\usepackage{soul}
\usepackage{todonotes}
\usepackage{longtable}
\usepackage{array}
\usepackage{pdflscape}
\usepackage[hypcap=false]{caption}
\usepackage{changepage} 

\usepackage[export]{adjustbox}
\usepackage{gensymb}

\newcommand{\allsdss}{SDSS~I~to~IV}
\newcommand {\NCVs}{504 }
\newcommand {\Nnew}{61 }
\newcommand {\Nspec}{248 }

\newcommand {\Nnewperiods}{82 }

\newcommand {\NmiscatCVs}{13 }
\newcommand {\scottsample}{366 }
\newcommand {\scottrecovered}{329 }
\newcommand {\foundspectra}{776 }
\newcommand {\foundbyscott}{21 }
\newcommand {\foundfromcvcandidatescarton}{6 }

\newcommand {\platespectra}{302 }
\newcommand {\platefound}{254 }
\newcommand {\plateunseen}{184 }
\newcommand {\platefp}{439 }
\newcommand {\sdallspectra}{687 }
\newcommand {\sdallCVs}{507 }
\newcommand {\sdallfound}{616 }
\newcommand {\sdallunseen}{180 }
\newcommand {\sdfoundCVs}{471 }
\newcommand {\Spectrafromcartons}{685 }
\newcommand {\CVsfromcartons}{422 }
\newcommand {\Newspectra}{431 }
\newcommand {\CVsnotfromcartons}{55 }
\newcommand {\notfoundspectra}{27 }
\newcommand {\sdfromCVcartons}{117\,841 }
\newcommand {\CVcartonspectra}{33\,064 }
\newcommand {\Veffugemten}{202}
\newcommand {\Veffugemsixty}{11}
\newcommand {\Veffsuumaten}{131}
\newcommand {\Veffsuumasixty}{56}

\newcommand {\spectrainsdssV}{2\,008\,677}

\newcommand {\SDSSV}{SDSS\nobreakdash-V}
\newcommand{\code}[1]{
        \begin{adjustwidth}{2cm}{}
        \texttt{#1}
        \end{adjustwidth}}

\newcommand\jaavso{AAVSO} 

\usepackage{newtxtext,newtxmath}

\title[CVs from SDSS V]{Cataclysmic variables from Sloan Digital Sky Survey\,-\,V (2020-2023) identified using machine learning. }

\author[K. Inight et al.]{
Keith~Inight,$^{1}$\thanks{E-mail: keith.inight@gmail.com }
Boris~T.~G\"ansicke,$^{1}$
Axel~Schwope,$^{2}$
Scott~F.~Anderson,$^{3}$
Elm\'{e}~ Breedt,$^{13}$
\newauthor
Joel~R.~Brownstein,$^{6}$
Sebastian~Demasi,$^{3}$
Susanne~Friedrich,$^{10}$
J.~J.~Hermes,$^{7}$
\newauthor
Knox~S.~Long,$^{9}$
Timothy~Mulvany,$^{1}$
Gautham~Adamane~Pallathadka,$^{8}$
Mara~Salvato,$^{10}$
\newauthor
Simone~Scaringi,$^{4,14}$
Matthias~R.~Schreiber,$^{5}$
Guy~S.~Stringfellow,$^{11}$
John~R.~Thorstensen,$^{12}$
\newauthor
Gagik~Tovmassian,$^{15}$
Nadia~L.~Zakamska$^{8}$
\\
$^{1}$Department of Physics, University of Warwick, Coventry, CV4 7AL, UK\\
$^{2}$Leibniz-Institut für Astrophysik Potsdam (AIP), An der Sternwarte 16, 14482 Potsdam, Germany \\
$^{3}$Astronomy Department, Box 351580, University of Washington, Seattle, WA 98195, USA \\
$^{4}$Centre for Extragalactic Astronomy, Department of Physics, Durham University, Durham, UK \\
$^{5}$Departamento de F{\'i}sica, Universidad T{\'e}cnica Federico Santa Mar{\'i}a, Avenida Espa{\~n}a 1680, Valpara{\'i}so, Chile\\
$^{6}$Department of Physics and Astronomy, University of Utah, Salt Lake City, UT 84112, USA\\
$^{7}$Department of Astronomy \& Institute for Astrophysical Research, Boston University, 725 Commonwealth Ave., Boston, MA 02215, USA\\
$^{8}$Department of Physics \& Astronomy, Johns Hopkins University, Baltimore, MD 21218, USA\\
$^{9}$Space Telescope Science Institute, 3700 San Martin Drive, Baltimore, MD, 21218, USA\\
$^{10}$Max-Planck-Institut für Extraterrestrische Physik, Giessenbachstraße, 85748 Garching, Germany\\
$^{11}$Center for Astrophysics and Space Astronomy, Department of Astrophysical and Planetary Sciences, University of Colorado, Boulder, CO, 80309, USA\\
$^{12}$Department of Physics and Astronomy, Dartmouth College, Hanover NH 03755, USA \\
$^{13}$Institute of Astronomy, University of Cambridge, Madingley Road, Cambridge CB3 0HA, UK \\
$^{14}$INAF-Osservatorio Astronomico di Capodimonte, Salita Moiariello 16, I-80131 Naples, Italy\\
$^{15}$Universidad Nacional Aut\'onoma de M\'exico, Instituto de Astronom\'{i}a, Aptdo Postal 106, Ensenada 22860, Baja California, M\'exico\\
}

\date{Accepted XXX. Received YYY; in original form ZZZ}

\pubyear{2020}

\begin{document}
\label{firstpage}
\pagerange{\pageref{firstpage}--\pageref{lastpage}}
\maketitle

\begin{abstract}
\SDSSV\ is carrying out a dedicated survey for white dwarfs, single and in binaries, and we report the analysis of the spectroscopy of \NCVs cataclysmic variables (CVs) and CV candidates obtained during the first 34 months of observations of \SDSSV. We developed a convolutional neural network (CNN) to aid with the identification of CV candidates among the over 2\,million \SDSSV\ spectra obtained with the BOSS spectrograph. The CNN reduced the number of spectra that required visual inspection to $\simeq2$~per cent of the total. We identified \foundspectra\ CV spectra among the CNN-selected candidates, plus an additional \notfoundspectra\ CV spectra that the CNN misclassified, but that were found serendipitously by human inspection of the data. Analysing the \SDSSV\ spectroscopy and ancillary data of the \NCVs\ CVs in our sample, we report \Nnew\ new CVs, spectroscopically confirm \Nspec and refute \NmiscatCVs\ published CV candidates, and we report \Nnewperiods\ new or improved orbital periods. We discuss the completeness and possible selection biases of the machine learning methodology, as well as the effectiveness of targeting CV candidates within \SDSSV. Finally, we re-assess the space density of CVs, and find $1.2\times 10^{-5}\,\mathrm{pc^{-3}}$. 
\end{abstract}



\section{Introduction}
Cataclysmic variables (CVs) are close binaries consisting of a white dwarf accreting from a donor star that is filling its Roche lobe \citep{2003cvs..book.....W}. CVs form an invaluable laboratory for testing our understanding of close binary evolution and accretion physics \citep{2020AdSpR..66.1004H} and the effects of strong white dwarf magnetic fields \citep{1990SSRv...54..195C,2015SSRv..191..111F,2021NatAs...5..648S}. CVs with very short periods ($<65$\,min) also generate measurable amounts of gravitational wave radiation which will be used to verify the performance of the \textit{LISA} space mission \citep{2018MNRAS.480..302K,2023MNRAS.525L..50S}.

There are many open questions regarding the evolution of CVs. One example is how longer-period CVs lose angular momentum, which is thought to be due to magnetic braking, although the details and rate of mass loss from the donor are still subject to debate \citep{2024A&A...682A..33B,2020MNRAS.491.5717B,2022MNRAS.517.4916E,2010A&A...513L...7S,2011ApJS..194...28K}. A second example is the lack of post period-minimum ``period-bouncers'' (\citealt{2023MNRAS.525.3597I} and references therein) and a further example is how the late emergence of magnetic fields affects the evolutionary path of CVs \citep{2021NatAs...5..648S}.

Complete and bias-free volumetric censuses of CVs are needed to validate evolutionary models \citep{2021MNRAS.504.2420I}. However, the only complete survey to date \citep{2020MNRAS.494.3799P} is limited to $150$\,pc and suffers from a small sample size (only 42 CVs). The  $300$\,pc sample of \citet{2021MNRAS.504.2420I} claims a completeness of $\sim50$ per cent. Larger volumetric censuses are therefore required, but achieving completeness is difficult due to large selection biases in the conventional methods used to identify CVs. In particular, whilst optical transients have revealed many CVs, they cannot detect non-outbursting CVs such as polar and novalike CVs \citep{2005ASPC..330....3G}. 

The Sloan Digital Sky Survey  (SDSS -- \citealt{2000AJ....120.1579Y,2006AJ....131.2332G}) has revolutionised the process of identifying and classifying CVs by providing large-scale spectroscopic data. These spectra have yielded many short-period and non-outbursting CVs \citep{2002AJ....123..430S,2003AJ....126.1499S,2004AJ....128.1882S,2005AJ....129.2386S,2006AJ....131..973S,2007AJ....134..185S,2009AJ....137.4011S,2011AJ....142..181S,2023MNRAS.524.4867I,2023MNRAS.525.3597I} and helped to address the selection biases within the historic CV samples \citep{2009MNRAS.397.2170G}. All of these spectra were manually identified using visual inspection. Following 20 years of using hard-wired optical fibre configurations called ``plug plates''  \citep{2003AJ....125.2276B}, \SDSSV\,  \citep{2017arXiv171103234K} now uses a robotic fibre positioning system \citep{2020SPIE11447E..8OJ}\,and the fibre-fed optical BOSS spectrograph \citep{2013AJ....146...32S} to capture up to 500 spectra simultaneously resulting in the \spectrainsdssV\,spectra\footnote{In more detail, the data set we used is the full set of \SDSSV\,spectra reduced with the BOSS pipeline version v6\_1\_0, excluding 1\,191\,969 spectra where the robot was placed to feed the infrared APOGEE spectrograph. Those fibres record in most cases a blank sky spectrum, or, rarely, that of a random object falling into the fibre. These ``serendipituous'' BOSS spectra are easily identified by their file name structure, spec-\{FIELD\}-\{MJD\}-\{RA\}\{DEC\}.fits} obtained since the beginning of \SDSSV\ on 2020 October 24 up to 2023 September 9. 

\SDSSV\ was also the first spectroscopic survey to specifically include CV candidates as targets \citep{2023MNRAS.525.3597I}~--~accounting for \sdfromCVcartons\ spectra out of the total \spectrainsdssV. This has proved invaluable, both for confirming existing CV candidates and also uncovering new CVs targeted because of an ultraviolet excess. However we recall that CVs discovered by SDSS in the first 20 years were serendipitously identified while searching for other types of objects, primarily quasars. A comprehensive identification of CVs therefore requires an analysis of the entire set of spectra in \SDSSV. A visual inspection of all spectra would be a time-consuming and error-prone exercise. We therefore developed an automated approach to reduce the number of spectra requiring expert review. This approach will also be useful for analysing other, even more prolific, spectroscopic surveys such as DESI \citep{2016arXiv161100036D, 2016arXiv161100037D}, WEAVE \citep{2012SPIE.8446E..0PD} and 4MOST \citep{2016SPIE.9908E..1OD}.

\section{Machine learning to classify large survey spectroscopy}\label{sec:method}
We describe here our approach for shortlisting the spectra of potential CVs. Once a shortlist has been generated, we visually inspect the spectra and obtain orbital periods using the methods described in \citet{2023MNRAS.525.3597I}. We therefore do not discuss details on determining classifications and orbital periods any further here and focus upon the initial generation of shortlists.

Machine learning has been used to find CVs from survey data before. \citet{2022MNRAS.517.3362M, 2024MNRAS.527.8633M} used machine learning to analyse photometry from \textit{Gaia} Alerts and ZTF, respectively, yielding $\simeq2\,800$ CVs candidates from $13\,280$ \textit{Gaia} Alerts.  \citet{2020AJ....159...43H} analysed specific emission and absorption lines in several million  LAMOST spectra and found 380 CV spectra. \citet{2023MNRAS.521..760V} used  machine learning based upon neural network technology to identify and classify white dwarfs. More generally \citet{2022A&A...667A.144E,2024A&A...682A...5V} and \citet{2023A&A...679A.127G} used a binary machine learning classifier to analyse \textit{Gaia} spectra in the search for white dwarfs.

Our objectives are to not only find the maximum number of CVs in a survey but also estimate the degree of completeness (i.e. estimate how many were missed) and more importantly assess the degree of bias and hence whether some CV sub-types were systematically omitted. Our approach to shortlisting therefore emulates the traditional human process using machine learning (see \citealt{géron2022hands} for an introduction to machine learning).  A convolutional neural network (CNN) is used (see Appendix\,\ref{sec:appendix1} for details).  The CNN classifies whether a spectral observation is that of a CV or not solely on the properties of its spectrum. Whilst other facts could potentially be used, such as location on the HR diagram, such ancillary data is inevitably incomplete introducing complications and potential biases.

\begin{table}
\caption{\label{Table:training} Summary of the spectroscopic samples used to train and test the CNN. The initial samples were constructed as detailed in Sections,\,\ref{sec:test_train_cv}--\ref{sec:test_train_wd} and then randomly sub-sampled to create the final test/train dataset.  The test data is a randomly selected subset of 20 per cent of the final test/train dataset and the size of the test data is shown here for ease of comparison with Fig.\,\ref{fig:conf1}.\\$^*$ see Section\,\ref{sec:evaluationof CNN} for details.}
\begin{tabular}{llll}
\hline
                                         & Initial  & Final & Test \\ \hline
CVs                                            & 623               & 563*  & 113 \\
Galaxies                                       & 50\,000             & 1206  & \rdelim\}{6}{3mm}[1129] \tabularnewline
Quasars                                        & 16\,713             & 1200  \\
Stars                                          & 50\,000             & 842  \\
White dwarfs                                   & 20\,341             & 1200  \\
Detached white dwarf plus  & 1602              & 1198  \\ 
 \hspace{10mm}         main sequence binary\\
[0.5ex]
Total                                          &                   & 6209 \\ \hline
\end{tabular}
\end{table}

\subsection{Training / test data}
The effectiveness of our approach is largely governed by the quality of the data used for training and testing the CNN (hereafter the train/test dataset). For our CNN the train/test dataset consists simply of a list of spectra and labels that are either ``1'' or ``0'' according to whether or not a given spectrum is that of a CV. The spectra in the train/test dataset  are derived by firstly resampling onto a standard grid of 3456 regularly spaced wavelengths between  $3860-8554$\,\AA\,using the \textsc{python specutils fluxcon} method \citep{nicholas_earl_2023_10016569}. The spectra are then smoothed using a five-point \textsc{specutils boxsmooth} filter and finally flux-normalised by dividing by their mean flux. A signal-to-noise (S/N hereafter) ratio is also calculated using  \textsc{specutils snr} on each spectrum. A small proportion (less than one per cent) of spectra were rejected by the \textsc{specutils} software due to having too many data points that were either missing or flagged as errors. 

For our train/test dataset we therefore needed a reliable sample of spectra of known CVs together with a reliable sample of non-CV spectra. The non-CV spectra were obtained by amalgamating samples of the following five categories: galaxies, quasars, stars, white dwarfs and detached white dwarf-main sequence binaries. Details on the individual samples are provided below and in Table\,\ref{Table:training}. The size of the non-CV sample was limited to the optimal number  (see Section\, \ref{sec:pnoncv}) of $\sim6000$~--~roughly 10 times that of the CV sample. In each case the Plate, MJD, and FIBER  attributes were used to download the spectra from SDSS DR17.

\subsubsection{CVs}
\label{sec:test_train_cv}
We chose to create our sample of 623 CV spectra using those described in \citet{2023MNRAS.524.4867I} and \citet{2023MNRAS.525.3597I} containing 507 and 118 CV spectra, respectively\footnote{34 CVs were reported in both papers. In these cases we included the published spectra  from both papers in our sample as  they used slightly different pipeline versions. Two spectra, spec-00614-53437-0251 and spec-15016-59146-4551094200 were omitted as they could not be resampled. }.  All of these have been recently studied and classified in a consistent manner. Furthermore the analysis in \citet{2023MNRAS.524.4867I} demonstrates a good coverage of the different sub-types. As the sample size (623) is relatively small the entire set was initially used without any filtering for quality. However, following some initial testing, we subsequently removed 60 spectra (see Section\,\ref{sec:evaluationof CNN}), hence the CV sample consists of 563 spectra.

\subsubsection{Galaxies}
We obtained our sample of galaxy spectra from SDSS DR17 \citep{2022ApJS..259...35A}, seeking the spectra of objects with reliable redshifts of $z>0.01$ and Petrosian magnitudes $g<18.5$ that had been categorised as galaxies by the SDSS pipeline. A set of 50\,000 spectra of galaxies was retrieved  using the following SQL query in CasJobs \citep{2008CSE....10...18L}:

\code{select top 50000 *\\
from specphotoall\\
into mydb.galaxy\_spectra\\
where type = 3\\
  and zWarning = 0\\
  and zErr < 1e-4\\
  and z > 0.01\\
  and petroMag\_g < 18.5}

\noindent
The galaxy spectra  were sorted by redshift ($z$) and a regular sample (every 41st spectrum) selected to get an even distribution of redshift resulting in a sample of 1206.

\subsubsection{Quasars}
We took our quasar sample from \citet{2003AJ....126.2579S}. Although there are later, larger catalogues we considered that this early work would be particularly reliable, having been subject to a number of visual examinations. The table was downloaded from \textsc{TapVizieR} \citep{2013ASPC..475..227L} using the following  statement:

\code{SELECT  "J/AJ/126/2579/table1".DR1, \\ "J/AJ/126/2579/table1".RAJ2000, \\     "J/AJ/126/2579/table1".DEJ2000,'\\
        '"J/AJ/126/2579/table1"."S-MJD", \\ "J/AJ/126/2579/table1".PNum, \\ "J/AJ/126/2579/table1".FNum \\FROM "J/AJ/126/2579/table1"'}

\noindent
We confirmed that there were no known CVs or candidates in the resulting list of 16\,713 spectra, from which we then obtained a random sub-sample of 1200 spectra. 
       
\subsubsection{Stars}
The spectra of stars vary markedly according to spectral type and it is therefore important to include a broad selection of spectral types in our sample. We obtained an initial set of 50\,000 spectra of stars from SDSS DR17 \citep{2022ApJS..259...35A} using the following SQL query in CasJobs: 
 
 \code{select top 50000 *\\
    from SpecPhotoAll\\
    into mydb.stellar\_spectra \\
    where type = 6\\
      and zWarning = 0 \\
      and zErr < 1e-4 \\
      and z < 0.01 \\
      and psfMag\_g < 18.5 \\
      and psfMag\_g > 16.5}

\noindent
The SDSS pipeline attributes a spectral type (based on the subClass field in the SpecPhotoAll table) to each of these stars. There were 34 different subClass values present and a random sample of 35 of each was selected. In some cases there were fewer than 35 for a given subClass and hence the stellar sample consists of only 842 spectra.

\subsubsection{White Dwarfs}
\label{sec:test_train_wd}

The spectra of white dwarfs vary markedly, not just as a function of their effective temperature and gravity but also due to the composition of their atmospheres. Although there are other, less common sub-types, we created a sample consisting of equal quantities of DA, DB and DC white dwarf spectra.

We initially obtained a set of 20\,341 reliably classified white dwarf spectra from \allsdss\, (Gentile Fusillo, private communication). A sub-sample of 1200 spectra was then created by randomly selecting 400 DA, 400 DB, and 400 DC white dwarfs. 

\subsubsection{Detached white dwarf plus main sequence binaries}
The spectra of these objects are distinctive not just because they are a blend of the fluxes from the white dwarf and the main sequence star but also because they potentially include narrow emission lines due to the irradiation of the main sequence star by the white dwarf.

Our initial set consisted of the spectra of 1602 objects defined by \citet{2010MNRAS.402..620R} which we  cross-matched with SDSS DR16  \citep{2020ApJS..249....3A} using the Vizier tables J/MNRAS/402/620/table5 and V/154/sdss16. A sub-sample of 1198 was then randomly selected from the 1602 
spectra for the train/test dataset.

\begin{figure} 
\centering
\includegraphics[width=\columnwidth]{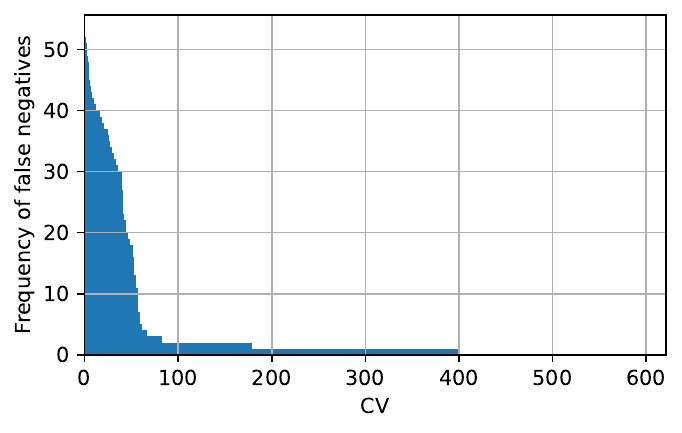}
\caption{\label{fig:Incidence of False Negatives} 
Frequency distribution of the individual CVs in the train/test sample (Table\,\ref{Table:training}) being identified as a false negative by the CNN,  from 200 random permutations of the sample. The horizontal axis represents the individual 623 CVs ordered by frequency. 220 CVs were always identified correctly and 90~per cent of the CVs occur at most three times as a false negative. 60 CVs frequently ($f>6$) resulted as false negatives indicating that they are not suitable for use in the training data (Section\,\ref{sec:test_train_cv}.)}
\end{figure}

\section{Testing / training the CNN} \label{sec:results}\label{section:testing}

\subsection{Training}
\label{sec:evaluationof CNN}

The CNN is trained by splitting the train/test dataset into two randomly chosen samples with 80~per cent being used for training and the remaining 20~per cent being tested to evaluate the result. To explore the effect of the random selection of the test sample we repeated the training 200 times with the selection of the training and test samples being determined randomly. Whilst the overall numbers of false positives and false negatives did not vary significantly it became apparent that certain CVs were consistently presenting as false negatives. We therefore analysed the set of false negatives in the 200 CNNs (see Fig.\,\ref{fig:Incidence of False Negatives}). From this it became apparent that 60 of the CVs were rarely if ever being found when included in the test sample. A review of the spectra of these 60 CVs revealed that 13 had significant quality issues (due to low S/N or a catastrophic failure in the pipeline), 11 exhibited cyclotron humps, 11 had relatively featureless blue novalike spectra, and nine were obtained during dwarf nova outbursts and resemble hot stars \footnote{ The remaining 16 were not exceptional.}. It is therefore not surprising that these 60 CV spectra (see Table\,3 in the supplementary data) were not correctly identified by the CNN. As these 60 CVs were not usefully contributing to the training data we removed them from the train/test dataset. The final CNN was then computed based on this revised train/test dataset of 563 CVs. The potential bias introduced by this is discussed further in Section\,\ref{sec:bias}.

The final CNN was trained using the train/test dataset listed in Table\,\ref{Table:training}. The training used 80 per cent of the CVs from the train/test database and 80 per cent of the non-CVs. The remaining 20 per cent of CVs and non-CVs were used to test the model (see final column in Table\,\ref{Table:training}.  We then computed a Confusion Matrix  for the final model (Fig.\,\ref{fig:conf1}) which shows that only six false positives were predicted and $110$ (out of a potential $113$) true  positives were found. 

\begin{figure}

\begin{tabular}{|ll|lc|}
\hline
\multicolumn{2}{|l|}{\multirow{2}{*}{}}                  & \multicolumn{2}{c|}{Predicted}                                                                                                                  \\ \cline{3-4} 
\multicolumn{2}{|l|}{}                                   & \multicolumn{1}{l|}{CV}                                                          & \multicolumn{1}{l|}{Not a CV}                                \\ \hline
\multicolumn{1}{|l|}{\multirow{2}{*}{Actual}} & CV       & \multicolumn{1}{c|}{\begin{tabular}[c]{@{}c@{}}110\\ (True positive)\end{tabular}} & \begin{tabular}[c]{@{}c@{}}3\\ (False negative)\end{tabular}   \\ \cline{2-4} 
\multicolumn{1}{|l|}{}                        & Not a CV & \multicolumn{1}{c|}{\begin{tabular}[c]{@{}c@{}}6\\ (False positive)\end{tabular}}  & \begin{tabular}[c]{@{}c@{}}1123\\ (True negative)\end{tabular} \\ \hline
\end{tabular}
 \caption{\label{fig:conf1} The confusion matrix resulting from training where the final CNN is applied to the 1242 test spectra (20 per cent) from the train/test dataset.  }
\end{figure}

\begin{table*}
\caption{\label{Table:numbers} A summary of the findings of the tests described below. This demonstrates that the CNN will recover  $(\mathrm{True\ positive/(True\ positive+False\  negative}))$ around $\simeq90$\,per cent of CVs   whilst reducing the sample to be manually reviewed $\mathrm{((True\ positive+False\ positive)/(Sample\ size}))$\ to $1-2$\,per cent.  }
 \centering
\begin{tabular}{llll}
\hline
\multirow{2}{*}{Test} & Sample size          & \multicolumn{2}{c}{Spectra}                              \\ \cline{3-4} 
                      & (spectra)            & True positive        & \multicolumn{1}{c}{False positive} \\ \hline
Plate (\ref{sec:platesurvey})                 & \CVcartonspectra                                               & \platefound           & \platefp             \\
CV candidate carton (\ref{sec:CVcandidates})  & 288                                                      & 200           &  0                          \\
The BOSS QA sample (\ref{sec:informal})        & \scottsample                          &                              \scottrecovered        &       0           \\
CV-related cartons (\ref{sec:CVtargsinSDSSV})   & \sdfromCVcartons                                                  & \Spectrafromcartons           & 1054                \\

All SDSSV (\ref{sec:allSDSSV})            & \spectrainsdssV\                       &                              \foundspectra&38\,402                    \\ \hline

\end{tabular}
\end{table*}

\subsection{Testing}

We tested the performance of the trained CNN using four largely independent samples of SDSS spectra, which had been human-classified to different levels of detail. 

\subsubsection{The plate survey of \texorpdfstring{\SDSSV}{xx}}\label{sec:platesurvey}

The \SDSSV\,targets are defined in a series of `cartons' each of which has a different science goal (see Table\,2 in \citealt{2023ApJS..267...44A}). \SDSSV\,included a number of CV-related cartons that were focused upon UV-excess objects and also catalogues of candidate CVs and white dwarfs.

A reliable set of 118 CVs were identified from the first eight months of \SDSSV\ (the ``plate'' survey,  \citealt{2023MNRAS.525.3597I}). There are \platespectra\ observations of these 118 CVs amongst the \CVcartonspectra BOSS spectra targeted by the CV-related cartons\footnote{\texttt{mwm\_wd\_core, mwm\_cb\_gaiagalex, mwm\_cb\_uvex1, mwm\_cb\_uvex2, mwm\_cb\_uvex3, mwm\_cb\_uvex4} and \texttt{mwm\_cb\_uvex5, mwm\_cb\_cvcandidates }} (see \citealt{2023ApJS..267...44A} section 9.2). We analysed these \CVcartonspectra spectra which were reduced with the  \SDSSV\,pipeline version 6\_1\_0 \citep{2023ApJS..267...44A} using our CNN resulting in \platefound\ out of \platespectra CV spectra (99 CVs out of 118 - 84 per cent) being recovered with only \platefp\ false positives (Table \,\ref{Table:numbers}). Whilst 118 spectra were used in the train/test dataset  the remaining \plateunseen\ were unseen data. The low level ($\simeq1$\,per cent) of false positives is reassuring.

\subsubsection{Previously known CV candidates}\label{sec:CVcandidates}
The \texttt{mwm\_cb\_cvcandidates} carton of \SDSSV\ was generated from lists of candidate or known CVs and was therefore expected to contain a higher proportion of CVs than other cartons. Our CNN identified 200 of the 288 spectra of targets in  \texttt{mwm\_cb\_cvcandidates} obtained by \SDSSV\   as CVs. A subsequent human review of the other 88 spectra found that 35 were CVs but with poor spectrum quality, 38 were not CVs (i.e the CNN correctly identified them as true negatives), and only 15 were spectra of actual CVs with good quality spectra. 
The CVs identified from this are included in Table\,1 in the supplementary data.

\subsubsection{The BOSS QA sample}\label{sec:informal}

The high-level quality of \SDSSV\ BOSS spectroscopy is monitored routinely including  quick-look visual inspections of the pipeline reductions for nearly all science spectra. Inspections are typically accomplished within a few days from when the spectra were taken. This aspect of BOSS quality assurance (QA) primarily aims at a prompt and high-level assessment of the overall quality of the spectra across each field/MJD combination. Each field/MJD combination yields several hundred science plus standard star spectra which are visually (albeit briefly) reviewed. No attempt is made in this QA to systematically verify the specific BOSS pipeline classifications for each individual target/spectrum. Whilst not their prime focus the QA team does, however, identify and record on an ad hoc basis particularly noteworthy or unusual individual spectra noticed in the course of this prompt high-level QA, including some potential CV candidates. 

We reviewed the sample of candidate CV spectra provided by the QA team, removed non-CVs and used the remainder as an independent check on the completeness of our CNN. Our CNN identified \scottrecovered\ out of the \scottsample\ (90 per cent) confirmed CV spectra identified by the QA team. It is important to note that in some cases the CNN will have detected the CV from a different spectrum and hence the recovery rate will be slightly understated.

\subsubsection{All spectra in \allsdss}\label{sec:sdssItoIV}
The vast majority of CVs identified among the spectra obtained within \allsdss\ were found via visual inspection  (\citealt{2002AJ....123..430S,2003AJ....126.1499S,2004AJ....128.1882S,2005AJ....129.2386S,2006AJ....131..973S,2007AJ....134..185S,2009AJ....137.4011S,2011AJ....142..181S}, see \citealt{2023MNRAS.524.4867I} for a detailed overview). A grand total of \sdallCVs\ CVs (\sdallspectra\ spectra) for which SDSS spectroscopy is available, with classifications at the sub-type level for most of them \citep{2023MNRAS.524.4867I}, provides the largest sample to test the CNN. We analysed all 5\,764\,653 BOSS spectra obtained by \allsdss\ with our CNN, which recovered \sdallfound\ out of  \sdallspectra\ spectra representing \sdfoundCVs\ (93 per cent) of the \sdallCVs\  known CVs (from \citealt{2023MNRAS.524.4867I}). Whilst \sdallCVs\ spectra were used for training,  \sdallunseen\  spectra were unseen by the CNN.

\section{Global application of the CNN}

\subsection{CV-related cartons in \texorpdfstring{\SDSSV}{xx}}\label{sec:CVtargsinSDSSV}

Up to $\mathrm{MJD}=60196$, \SDSSV\ obtained \sdfromCVcartons\ BOSS spectra of targets within one or more CV-related cartons. This sample included 295 spectra of previously known CVs plus an unknown number of new  CVs.  As expected our CNN recovered $254$ out of the $295$ CV spectra and found an additional 1488 candidate spectra. An analysis of the 19 CVs (41 spectra) that were not recovered showed that 14 had severe flux calibration issues, whilst the remainder had indistinct emission lines.  The 1488 spectra were manually examined and \Newspectra were found to be CV spectra yielding a total of \Spectrafromcartons\  true positives (\CVsfromcartons\ CVs). 

\subsection{All targets in \texorpdfstring{\SDSSV}{xx}}\label{sec:allSDSSV}

Recalling that some CVs in \SDSSV\,will only be found in unrelated cartons, we analysed the  entire \spectrainsdssV\,spectra in \SDSSV\,up to $\mathrm{MJD}=60196$ using our CNN resulting in 39\,181 spectra ($\simeq2$\,per cent of the total) being classified as candidate CVs. These spectra were examined by eye \footnote{The co-authors reviewed randomly selected subsets of the 39\,181 and then the author separately reviewed the entire set to ensure consistency. The process was facilitated by some bespoke \textsc{python} code so that each spectrum could be classified using a single keystroke.} for CV-related features yielding \foundspectra true positives and 38\,402 false positives. The \foundspectra spectra include \CVsnotfromcartons  CVs that were not in the CV-related cartons.

\subsection{Biases and CNN performance}

\subsubsection{False positives}

To characterise the false positives arising from the CNN we analysed the 1488 spectra identified in Section\,\ref{sec:CVtargsinSDSSV}. 1054 of these spectra proved not to be CVs. Of these 1054 spectra 99 were catastrophic failures with random appearance. All but three of the remaining 955 spectra exhibited some kind of emission lines. These were all thin lines unlike those of CVs. 474 of these 955 included \ion{H}{} or \ion{He}{} emission lines.  

In passing we note that some objects identified by the CNN are interesting even though they are not CVs. One example is SDSS\,J102347.69+003840.5 (AY\,Sex) which is a low mass X-ray binary where the compact object is actually a neutron star \citep{2005AJ....130..759T,2009Sci...324.1411A,2019MNRAS.488..198S,2013arXiv1311.5161A}.

\subsubsection{Does the S/N of spectra affect the reliability of the CNN?}
\begin{figure} 
\centering
\includegraphics[width=\columnwidth]{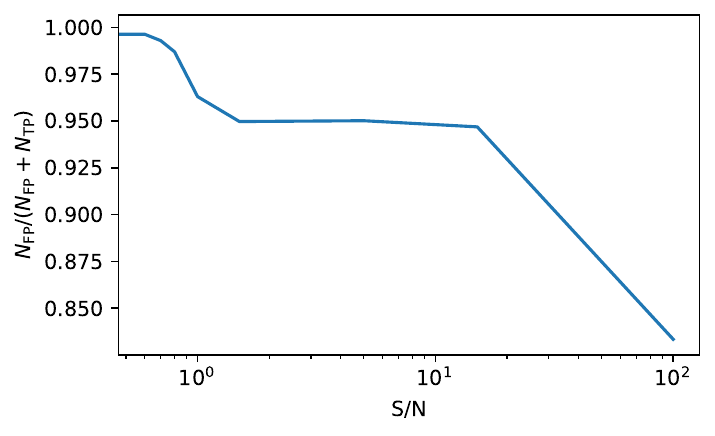}
\caption{\label{fig:SNR} 
The fraction of false positives with respect to the sum of false plus true positives as a function of S/N based on the 39\,181 CV candidates that the CNN identified among the total \spectrainsdssV\ spectra that were analysed. The proportion of false positives is relatively flat, $\simeq0.95$, until $\mathrm{S/N}\lesssim1$. Nevertheless, even at very low S/N some true positives are identified. The relatively small number of spectra with $\mathrm{S/N} \gtrapprox 15$ yield relatively few false positives.}
\end{figure}

Before discussing selection biases of individual CV sub-types we consider whether low S/N systematically affects the outcome of our CNN analysis\footnote{It is important to note that there are reasons other than low S/N for a poor quality spectrum, including poor flux calibration, partially missing spectral coverage, or strong sky line residuals}. We analysed the effect of S/N by considering the fraction of false positives in the sample of \spectrainsdssV\  spectra in \SDSSV\ up to $\mathrm{MJD}=60196$ (Fig.\,\ref{fig:SNR}). To assess the false negatives was impractical as it would require the visual analysis of \textit{all} \spectrainsdssV\ spectra and we assume that the behaviour of the false negative rate is similar to that of the false positive rate. The proportion of false positives ($\simeq0.95$)  is relatively flat for $1<\mathrm{S/N}<10)$. 

\begin{figure} 
\centering
\includegraphics[width=\columnwidth]{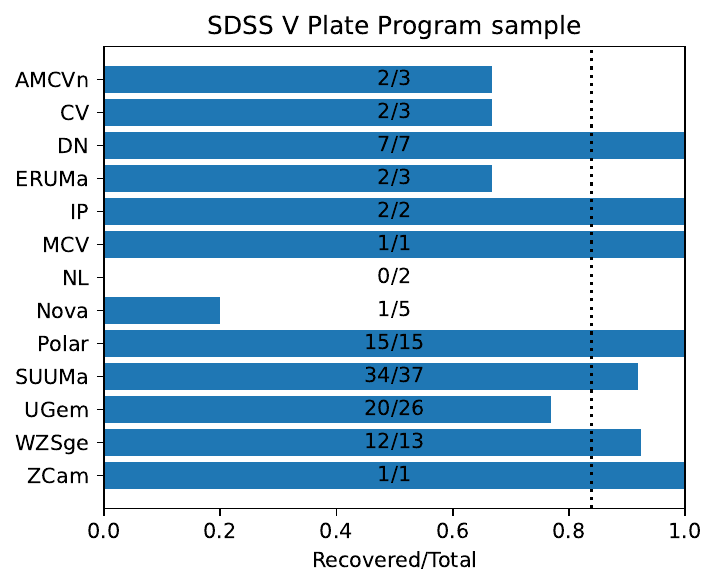}
\includegraphics[width=\columnwidth]{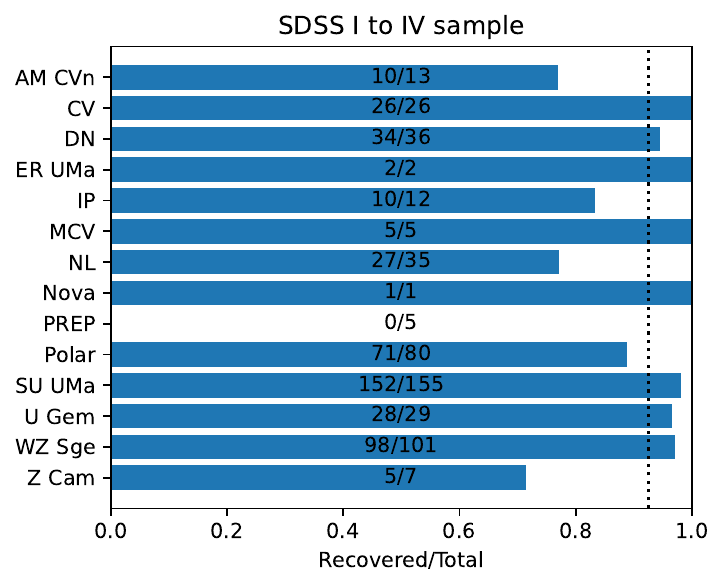}
\includegraphics[width=\columnwidth]{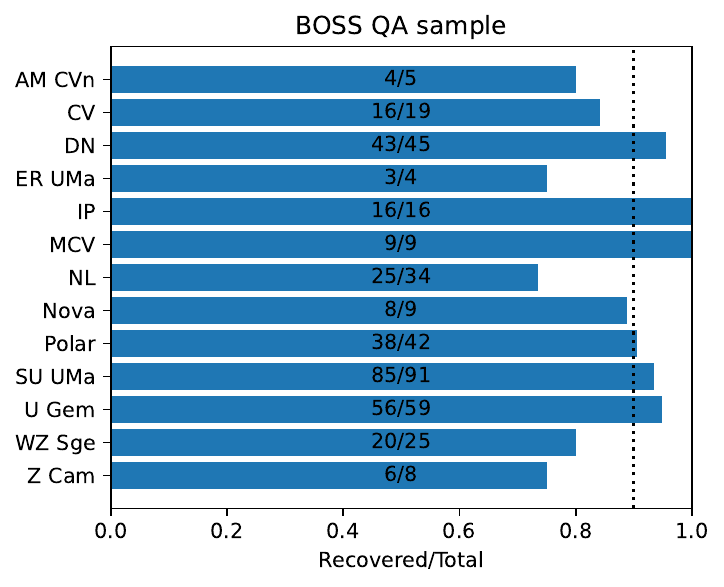}
\caption{\label{fig:Bias} 
Machine learning performance broken down by CV sub-type (as defined in Section\,3 in \citealt{2023MNRAS.524.4867I}) where  MCV\,=\,Magnetic CV, PREP\,=\,pre-polar, IP\,=\,Intermediate Polar. The fraction of CVs found (=\,true positives) by the CNN is shown for the \SDSSV\ sample (top panel), the \allsdss\ sample (middle panel) and the BOSS QA sample (bottom panel). The overall recovery rate is shown as a dotted line in each case.  }
\end{figure}

\begin{figure} 
\centering
\includegraphics[width=\columnwidth]{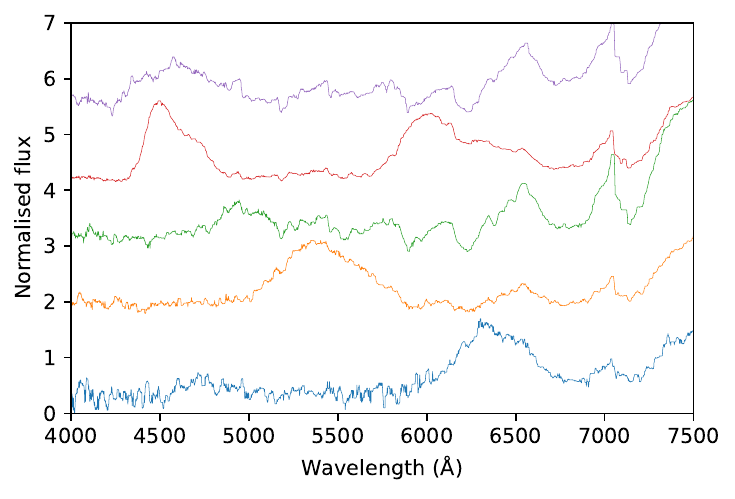}
\caption{\label{fig:preps} 
Spectra of pre-polars from \allsdss\,showing the characteristic cyclotron humps. From top to bottom: 
J204827.90+005008.9, J155331.11+551614.4, J120615.73+510047.0, J130147.97+434549.4,
J105905.06+272755.4. None of these five systems were correctly identified as CVs by the CNN (Section\,\ref{sec:bias}).}
\end{figure}

\subsubsection{CV sub-types and potential selection bias}\label{sec:bias}
The potential biases of the CNN would ideally be assessed using an unseen (by the CNN) set of CVs which have been classified by sub-type and which also have SDSS spectra. Unfortunately such a dataset does not exist as all SDSS CVs known prior to this project have already been used in the train/test dataset. We therefore used three non-ideal methods with the assumption that, combined, they will reveal any significant biases.   

We considered first the fractions of each CV sub-type within the sample of 118 CVs published in \citet{2023MNRAS.525.3597I} that were recovered by the CNN (top panel of Fig.\,\ref{fig:Bias}). These 118 CVs were of course part of the train/test dataset used to train the CNN but, with this caveat in mind, it can be seen that the overall recovery rate does not significantly depend upon sub-type except for poor performance with classical novae and novalikes. One reason for the overall poor recovery rate is that \SDSSV\ spectra can be distorted by incorrect flux calibration, explaining 14 of the 19 CVs that the CNN classified as false negatives. In addition, the sample size of the \SDSSV\ plate-program CVs  is also small.

Secondly we considered the \allsdss\ sample. This data was also used to train the CNN. However it is a much larger sample and the \allsdss\ spectra do not not suffer from poor flux calibration. This analysis (middle panel of Fig.\,\ref{fig:Bias}) shows relatively little bias with the exception of pre-polars, with none of the five known systems being recovered. Pre-polars are a rare sub-type of CV, and their spectra are dominated by cyclotron humps (see Fig.\,\ref{fig:preps}). The shape, size and wavelength of these humps is a strong function of the magnetic field strength, which, coupled with their rare nature, probably means that there is insufficient training data and hence it is hard, if not impossible, for our CNN to recognise pre-polars.

Thirdly we considered the BOSS QA sample. This is essentially an unseen sample and shows very little bias.  The overall recovery rate for the  BOSS QA sample is 90~per cent.

In summary, with the exception of pre-polars,  the three samples show a consistent recovery rate with little apparent bias between sub-types.
\clearpage

\onecolumn
\begin{landscape}
\begin{longtable}[c]{lllllrllll}
\caption{\label{tab:cvlist} CVs from \SDSSV. We provide the geometric distances from \citet{2021AJ....161..147B} and CV sub-types according to the definitions in Section\,3 in \citet{2023MNRAS.524.4867I}.   The references for the initial discovery (ID), spectrum (Sp)  and orbital and superhump periods ($P$~--~where known) are also shown. New discoveries, orbital periods and sub-type identifications are shown in blue. Sub-type identifications in black are from VSX \citep{2017yCat....102027W}. : Tentative values, * Superhump periods  }\\
\hline
\multirow{2}{*}{SDSS} & \multirow{2}{*}{Name} & \multirow{2}{*}{\begin{tabular}[c]{@{}l@{}}Gaia EDR3\\ source\_id\end{tabular}} & \multirow{2}{*}{\begin{tabular}[c]{@{}l@{}}Period \\ (h)\end{tabular}} & \multirow{2}{*}{\begin{tabular}[c]{@{}l@{}}Gaia EDR3\\$G$ (mag)\end{tabular}} & \multirow{2}{*}{\begin{tabular}[c]{@{}l@{}}Distance\\ (pc)\end{tabular}} & \multirow{2}{*}{\begin{tabular}[c]{@{}l@{}}Variable\\ Sub-type\end{tabular}}  & \multicolumn{3}{c}{References}                         \\ \cline{8-10} 

 &    &     &       &      &      &         & \multicolumn{1}{l}{ID} & \multicolumn{1}{l}{Sp} & $P$ \\ \hline
 \hline
 \endfirsthead
\hline
\multirow{2}{*}{SDSS} & \multirow{2}{*}{Name} & \multirow{2}{*}{\begin{tabular}[c]{@{}l@{}}Gaia EDR3\\ source\_id\end{tabular}} & \multirow{2}{*}{\begin{tabular}[c]{@{}l@{}}Period \\ (h)\end{tabular}} & \multirow{2}{*}{\begin{tabular}[c]{@{}l@{}}SDSS g\\ (mag)\end{tabular}} & \multirow{2}{*}{\begin{tabular}[c]{@{}l@{}}Distance\\ (pc)\end{tabular}} & \multirow{2}{*}{\begin{tabular}[c]{@{}l@{}}Variable\\ Sub-type\end{tabular}} &  \multicolumn{3}{c}{References}                         \\ \cline{8-10} 
 &    &     &       &      &      &         & \multicolumn{1}{l}{ID} &\multicolumn{1}{l}{Sp} & P \\ \hline
\hline
\endhead
\hline
\endfoot
J000155.14--670743.3 & EF Tuc & 4707482485122305536 & 3.5: & 14.85 & $1329$ & NL & 14 & 13 & 24 \\
J000600.15+012129.8 &  & 2738755406045571968 & 1.52(10) & 19.57 & $352$ & WZ Sge: & 15 & 15 & 15 \\
J000633.70--690033.8 & HV 8001 & 4706297108508032128 & 1.90(3) & 15.1 & $241$ & SU UMa & 9 & 17 & 17 \\
J000844.33--014014.8 & Gaia15abi & 2544817841421994624 &  & 18.15 & $1236$ & NL:/Polar & 2 & 26 &  \\
J001204.50+020129.6 &  & 2546760133008036096 & \textcolor{blue}{3.288(1)} & 16.45 & $1667$ & \textcolor{blue}{NL} & 21 & 18 & 1 \\
J002243.55+061002.7 & MGAB-V295 & 2747622177049417344 & 1.9(1): & 19.61 & $2113$ & SU UMa & 30 & 26 & 26 \\
J002257.65+614107.5 & V1033 Cas & 430446643235641472 & 4.033 & 16.58 & $1557$ & IP & 5 & 10 & 10 \\
J002728.02--010828.7 & EN Cet & 2541910801397761152 & 1.424(1) & 20.76 & $1218$ & WZ Sge & 12 & 27 & 6 \\
J002848.83+591722.0 & V709 Cas & 428101969045877504 & 5.33290(7) & 14.39 & $725$ & IP & 8 & 8 & 11 \\
\textcolor{blue}{J003529.47--735826.3} &  & 4685792041310761984 &  &  & $957$ & \textcolor{blue}{CV:} & 1 & 1 &  \\
J003827.05+250925.0 & 1RXS J003828.7+250920 & 2806802123399581056 & 2.26826(2) & 18.63 & $503$ & SU UMa & 16 & 23 & 26 \\
J003941.08+005427.5 & SDSS J003941.06+005427.5 & 2543387617312121216 & 1.52(1) & 20.76 & $1078$ & WZ Sge: & 27 & 27 & 4 \\
J004335.16--003729.7 &  & 2530961280492678528 & 1.3721(14) & 19.85 & $447$ & WZ Sge: & 19 & 19 & 31 \\
J005000.03--760827.3 & BMAM-V424 & 4684361817175293440 & 3.698(4) & 14.9 & $1031$ & NL & 7 & 1 & 25 \\
J005050.87+000912.6 & GS Cet & 2537078211571341568 & 1.34 & 20.68 & $308$ & WZ Sge & 29 & 29 & 29 \\
J010411.62--031342.2 &  & 2531434796342548224 &  & 19.69 & $1096$ & SU UMa: & 20 & 3 &  \\
\textcolor{blue}{J012132.92--740819.5} &  & 4686117767317363072 &  & 16.91 & $2972$ & \textcolor{blue}{NL:} & 1 & 1 &  \\
J012754.58--581717.5 & ASASSN -15bg & 4909517162610311040 & 1.57(1)* & 20.43 & $730$ & SU UMa & 32 & 1 & 28 \\
\textcolor{blue}{J012905.53--580841.2} &  & 4909542996837627648 &  &  & $357$ & \textcolor{blue}{AM CVn} & 1 & 1 &  \\
J014227.07+001729.8 &  & 2510205490257050496 & 1.88(4) & 19.89 & $697$ & Polar: & 22 & 22 & 15
\end{longtable}
\noindent
This is a small sample of the data. The full table can be found in the supplementary data.\\
References:\, 1\,\textcolor{blue}{This work}, 2\,\citet{2021A&A...652A..76H}, 3\,\citet{2014AJ....148...63S}, 4\,\citet{2010A&A...524A..86S}, 5\,\citet{2006ATel..709....1H}, 6\,\citet{2008MNRAS.386.1568D}, 7\,\textcolor{blue}{AAVSO}, 8\,\citet{1996A&A...307..459M}, 9\,\citet{1933AN....249..395L}, 10\,\citet{2007A&A...473..185B}, 11\,\citet{2010PASP..122.1285T}, 12\,\citet{1997IAUC.6763....2E}, 13\,\citet{1995Ap&SS.230..101S}, 14\,\citet{2001MNRAS.325...89C}, 15\,\citet{2023MNRAS.525.3597I}, 16\,\textcolor{blue}{VSNET 12318}, 17\,\citet{2010MNRAS.405..621A}, 18\,\citet{2018yCat.5153....0L}, 19\,\citet{2004AJ....128.1882S}, 20\,\citet{2009ATel.2266....1D}, 21\,\textcolor{blue}{vsnet-chat 7946}, 22\,\citet{2017ApJS..228...19C}, 23\,\citet{2018AJ....155...28S}, 24\,\textcolor{blue}{CBA512}, 25\,\citet{2023MNRAS.520.3355S}, 26\,\citet{2023MNRAS.524.4867I}, 27\,\citet{2005AJ....129.2386S}, 28\,\citet{2015PASJ...67..105K}, 29\,\citet{2007MNRAS.382.1145S}, 30\,\textcolor{blue}{MGAB Variable Star Catalog}, 31\,\citet{2008MNRAS.391..591S}, 32\,\citet{2014ApJ...788...48S} 
\noindent
\end{landscape}
\twocolumn
\clearpage

\section{CVs found in \texorpdfstring{\SDSSV}{xx}} \label{subsec:CVtable}
The final collection of CVs observed by \SDSSV\ consists of the \CVsfromcartons from Section\,\ref{sec:CVtargsinSDSSV}, the \CVsnotfromcartons from Section\,\ref{sec:allSDSSV}, \foundbyscott CVs from the BOSS QA sample that were not found by the CNN and \foundfromcvcandidatescarton CVs from \texttt{mwm\_cb\_cvcandidates} that were also not found by the CNN.  These \NCVs\ CVs were further analysed following the methods in \citet{2023MNRAS.524.4867I,2023MNRAS.525.3597I}. This analysis of the \SDSSV spectroscopy and ancillary data of the \NCVs\ CVs in our sample yielded \Nnew\ new CVs, spectroscopic confirmation of \Nspec candidates and \Nnewperiods\ new or improved orbital periods. 
Prior to \SDSSV, one of the authors (JRT) obtained time-resolved spectroscopy using the MDM observatory at Kitt Peak (see \citealt{2020AJ....160....6T} for more details on the instrument and data reduction techniques) of a number of our CVs. We present here the so far unpublished periods resulting from these observations. The full set of data is included in the supplementary data and a sample is shown in Table\,\ref{tab:cvlist}. Whilst analysing the \NCVs\ CVs in \SDSSV\ some CVs with unusual or noteworthy characteristics were identified and these are discussed below. For simplicity the names of CVs have been abbreviated in the text by dropping the arcseconds; so for example SDSS\,J015543.47+002807.4 becomes J0155+0028.

\begin{figure} 
\centering
\includegraphics[width=0.83\columnwidth]{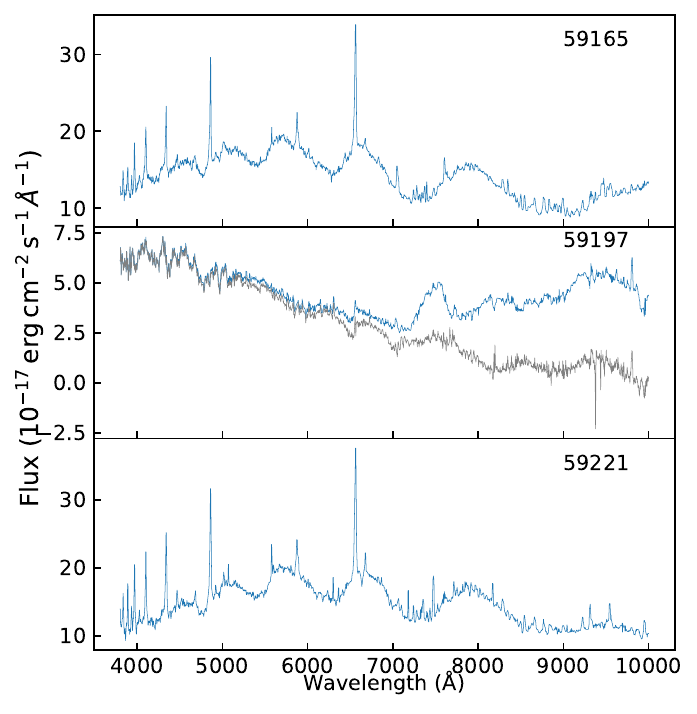}
\includegraphics[width=0.83\columnwidth]{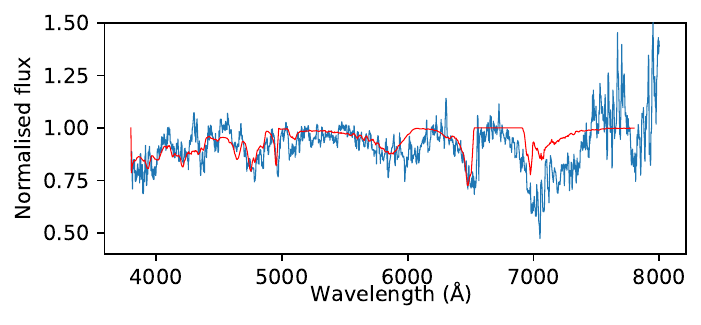}
\includegraphics[width=0.88\columnwidth]{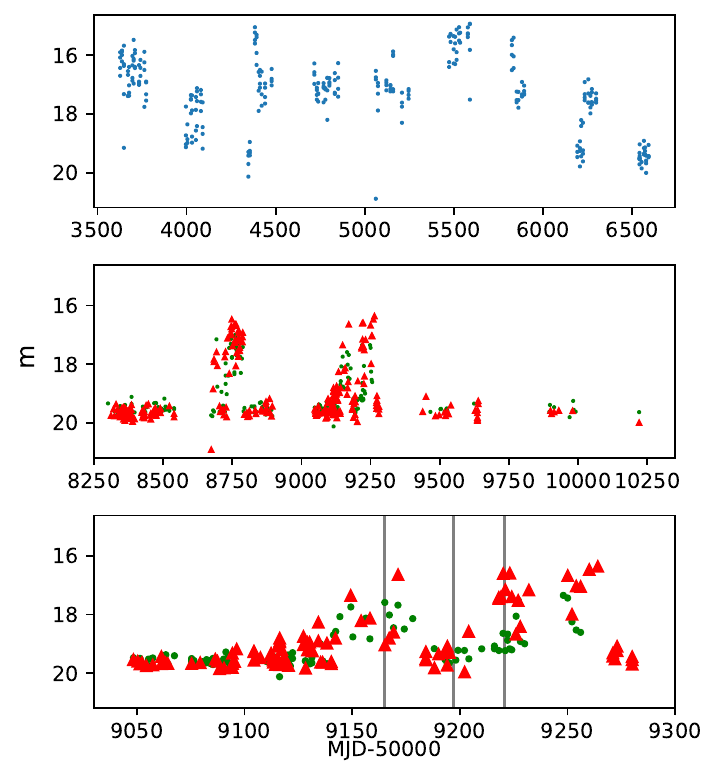}
\caption{\label{fig:J0155+0028}
High and low-state observations of the polar J0155+0028. The top three panels show the system observed twice in a high state ($\mathrm{MJD}=59165, 59221$ and once in a low state ($\mathrm{MJD}=59197$). The high-state spectra are dominated by cyclotron humps whilst the low-state spectrum shows a reduced flux level, no Balmer emission lines, and photospheric features of the white dwarf and the donor star. The grey line in the second panel shows the effect of subtracting the spectrum of an M7 dwarf. The fourth panel shows the normalised, donor-subtracted low-state spectrum (grey) and a model spectrum (red) of a slightly offset dipole with a polar field strength of 58\,MG (see text for details) which reproduces well the observed Zeeman-split H$\alpha$ to H$\delta$ line profiles. The CRTS light curve (fifth panel) shows erratic variability whilst the more recent ZTF light curve (sixth panel, $r$-band as red triangles, $g$-band as green dots) captured the system  mainly in a low state with two brief high states. The bottom panel shows a zoom-in of the ZTF light curve with the times of our three SDSS spectra indicated in grey.}
\end{figure}

\begin{figure} 
\centering
\includegraphics[width=\columnwidth]{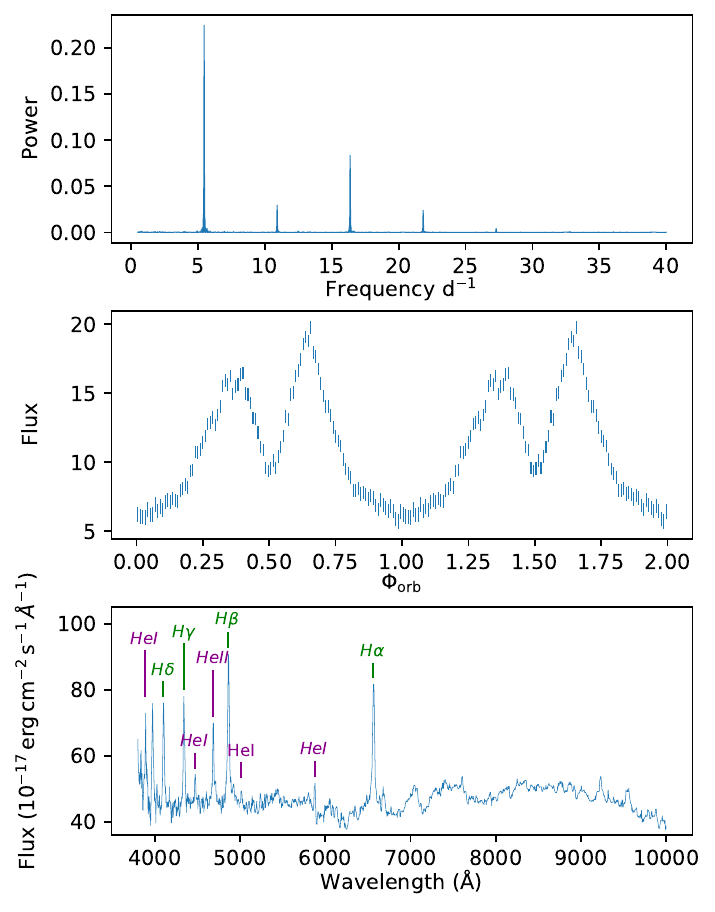}

\caption{\label{fig:J0649-0737}
Top panel: Lomb-Scargle periodogram computed from the \textit{TESS} light curve of J0649$-$0737. Middle panel: The phase-folded \textit{TESS} light curve on a period of $ 4.3988$\,h where $\phi=0$ is approximately at minimum flux. The flux is an uncertainty-weighted average of observations in $2.5$ min bins.  Bottom panel: SDSS spectrum of J0649$-$0737.   }
\end{figure}

\begin{figure} 
\centering
\includegraphics[width=\columnwidth]{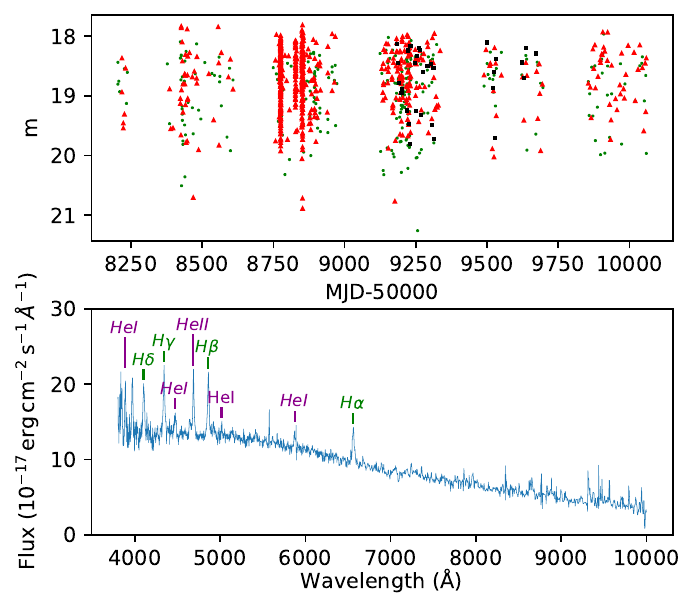}

\caption{\label{fig:J0804-0022} Top panel: The ZTF light curve of J0804--0022 with $r$-band data shown as red triangles, $g$-band data as green dots and $i$-band data as black squares. Bottom panel: SDSS spectrum of J0804$-$0022. Note the strength of the \ion{He}{II} emission line relative to H$\beta$.  }
\end{figure}

\begin{figure} 
\centering
\includegraphics[width=\columnwidth]{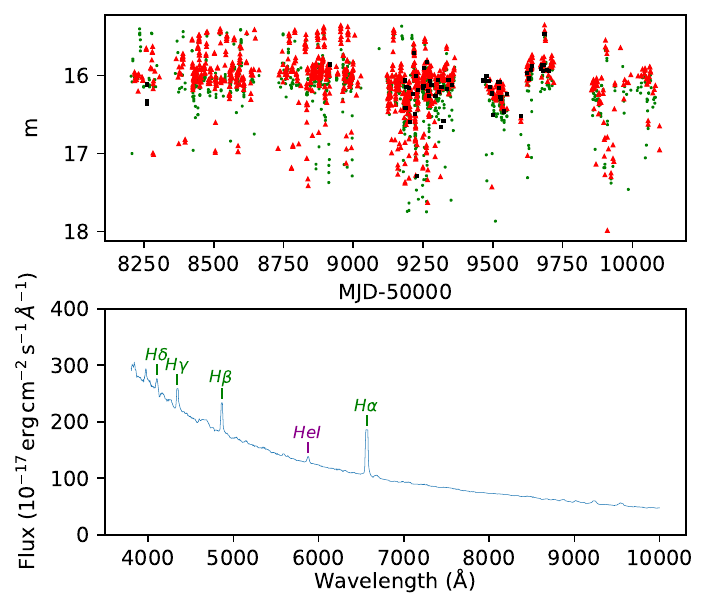}

\caption{\label{fig:J0907+7158}
Top panel: ZTF light curve of J0907+7158 with observations using the $r$-band filter shown in red triangles, $g$-band filter in green dots and $i$-band in black squares. The eclipses are evident and the pronounced variability outside the eclipses is unusual. Bottom panel: SDSS Spectrum of J0907+7158.  }
\end{figure}

\begin{figure} 
\centering
\includegraphics[width=\columnwidth]{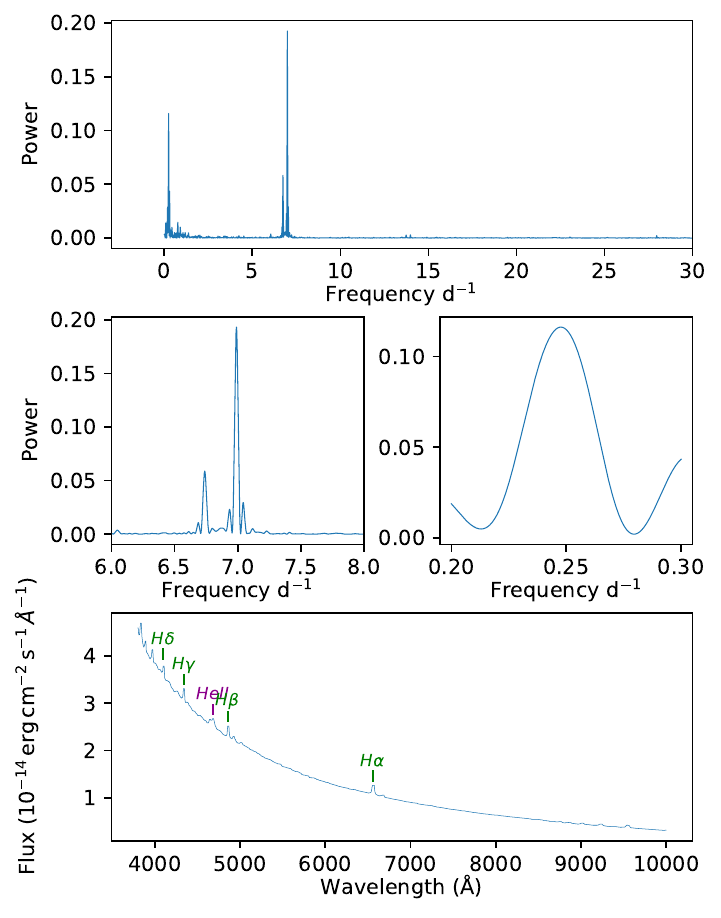}

\caption{\label{fig:J0925+4349}
Top panel: Periodogram from the \textit{TESS} light curve of J0925+4349. Centre panels: expanded plots showing the periods of $3.434$\,h ($6.99\,\mathrm{d^{-1}}$), $3.56$\,h ($6.74\, \mathrm{d^{-1}}$) and $4.04$\,d ($0.248\, \mathrm{d^{-1}}$). The peak at $6.99\, \mathrm{d^{-1}}$ is a beat between the orbital period ($6.99\, \mathrm{d^{-1}}$) and the precession period ($0.248\, \mathrm{d^{-1}}$). Bottom panel: SDSS Spectrum of J0925+4349. }
\end{figure}

\begin{figure} 
\centering
\includegraphics[width=\columnwidth]{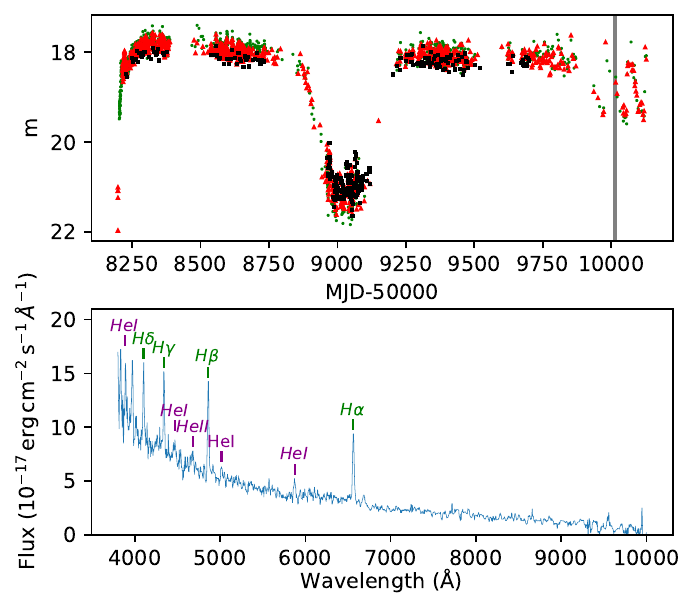}

\caption{\label{fig:J1604+4143}
Top panel: The ZTF light curve of J1604+4143 with the $r$-band filter shown as red triangles, $g$-band data as green dots and $i$-band data as black squares. The deep low state lasted six months followed by a steady high state lasting for two years before the onset of disc outbursts.  Bottom panel: The spectrum of J1604+4143 taken on \mbox{$\mathrm{MJD}=60114$}, indicated by the grey line in the top panel, during the decline from an outburst. }
\end{figure}

\begin{figure} 
\centering
\includegraphics[width=\columnwidth]{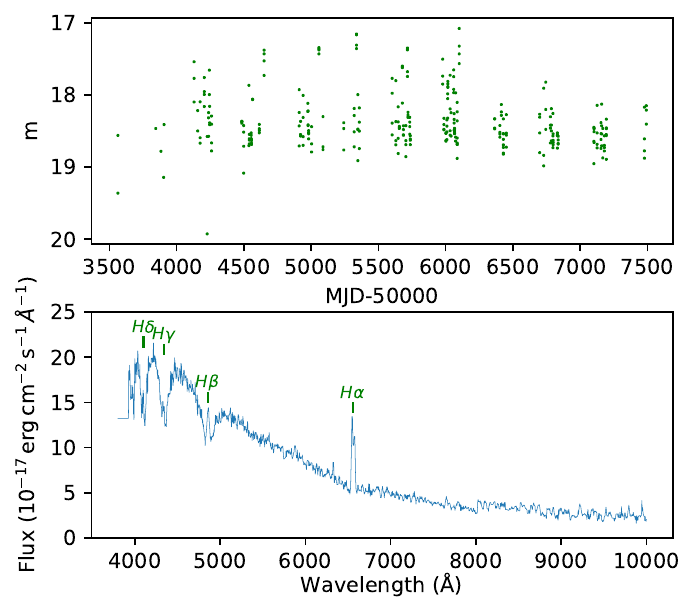}

\caption{\label{fig:J1608-1350}
Top panel: ZTF light curve ($g$-band filter) of J1608$-$1350 showing stunted outbursts atypical of short-period CVs. Bottom panel: Spectrum of J1608$-$1350 with the white dwarf signature clearly visible indicative of an old short-period CV. }
\end{figure}

\begin{figure} 
\centering
\includegraphics[width=\columnwidth]{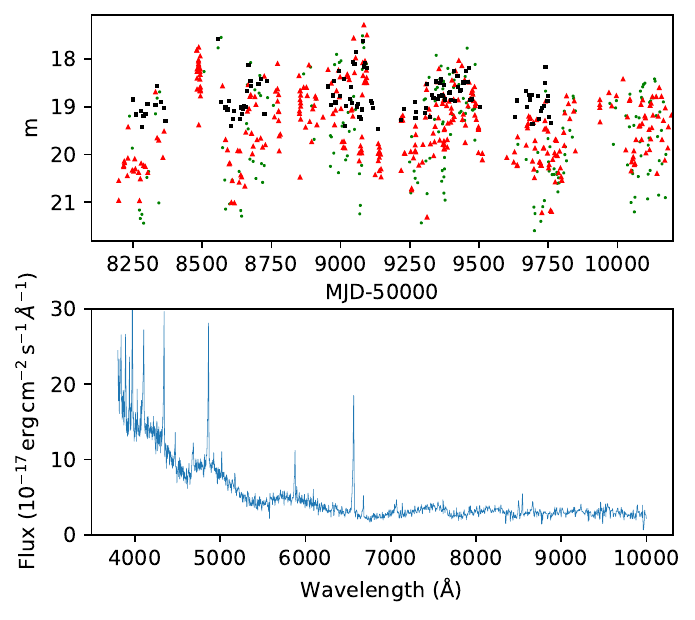}

\caption{\label{fig:J1618+0309}
Top panel: The ZTF light curve of J1618+0309 with the $r$-band filter shown as red triangles, $g$-band data as green dots and $i$-band data as black squares. There are clear high and low states with variability of $\Delta m\simeq1.5$ in the high state.   Bottom panel: The spectrum of J1618+0309 showing cyclotron humps.  }
\end{figure}

\begin{figure} 
\centering
\includegraphics[width=\columnwidth]{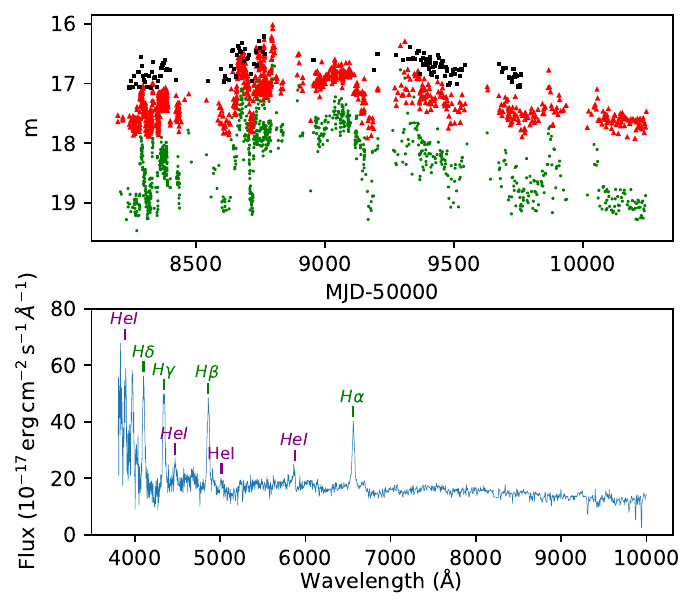}

\caption{\label{fig:J2007+6250}
Top panel: The ZTF light curve of J2007+6250 with the $r$-band filter shown as red triangles, $g$-band data as green dots and $i$-band data as black squares. Frequent state changes are evident. Bottom panel: The spectrum of J2007+6250 revealing a dominant contribution of the donor star. }
\end{figure}

\subsection{SDSS J015543.47+002807.4}

This system is also known as FL\,Cet. The SDSS spectra from $\mathrm{MJD}=59165$ and $\mathrm{MJD}=59221$ (Fig.\,\ref{fig:J0155+0028}) show pronounced cyclotron humps.  The spacing of the humps indicates a magnetic field of $\mathrm{B}\simeq23$\,MG, significantly lower than the  $B\simeq29$\,MG reported by \citet{2005ApJ...620..422S}. Zeeman split absorption lines of H$\alpha$ to H$\delta$ from the white dwarf photosphere are visible in the low-state spectrum obtained on $\mathrm{MJD}=59197$. The low-state spectrum also reveals the donor star, which is well-matched by an M-dwarf template with a spectral type of M7 (taken from \citealt{2017ApJS..230...16K}). 

The CRTS light curve (Fig.\,\ref{fig:J0155+0028}) captured the system mainly in a high state ($16-18$\,mag), in contrast, the more recent ZTF data shows the system predominantly in a low state ($\simeq20$\,mag), with two relatively short high states. The transitions between high and low states can occur on time scales of days, 

We modelled the donor-subtracted, normalised low-state spectrum using synthetic spectra computed with a semi-empirical method following \citet{1992A&A...257..353O}. A satisfactory fit to the data is achieved for a dipole slightly offset in $x-$ and $z-$direction by 0.03 and 0.05 stellar radii respectively, polar field strengths between 50\,MG and 58\,MG and the inclination angle between the line-of-sight and the magnetic axis of $75^\circ$. We only cursorily explored the large parameter space, other possible field configurations could provide similar good fits. In addition, our code does not use any radiative transfer calculation but uses the Planck function where the optical depth reaches unity, hence the resulting line profiles and depths are only approximate. Nevertheless, the wavelengths of the $\sigma$-components of H$\alpha$ at 5900\,\AA\ and 7000\,\AA\ require a field strength of at least 33\,MG.

\subsection{SDSS J064950.93--073741.0}

This system is also known as RX\,J0649.8-0737. 
\citet{2020MNRAS.491..201J} provides a detailed study of this polar. We obtained a \textit{TESS} light curve from which (Fig.\,\ref{fig:J0649-0737}) we derived a period of $4.3988(3)$\,h which we consider more accurate than the $4.347(1)$\,h in \citet{2020MNRAS.491..201J}. The binned folded light curve (Fig.\,\ref{fig:J0649-0737}) shows two peaks per orbital cycle, broadly resembling the $V$-band light curve of AM\,Her \citep{2001A&A...372..557G}. The orbital variation is therefore most likely due to cyclotron beaming. The SDSS spectrum of J0649--0737 shows a clear contribution of the donor star, as well as strong Balmer and He lines.

\subsection{SDSS J074222.55+172806.7}

This system is also known as MLS 110309:074223+172807. 
This object is eclipsing and we obtained a period of $2.47465(2)$\,h from the ZTF light curve which also shows outbursts.  This places the system within the period gap and it is likely to be an SU\,UMa dwarf nova, but in the absence of superoutbursts we only classify it as a dwarf nova.

\subsection{SDSS J075648.85--124653.9}

This system is also known as ATO\,J119.2035$-$12.7816.
The SDSS spectrum shows double-peaked Balmer and \ion{He}{i} emission lines. The object is on  the main sequence in the HR diagram just below the region occupied by \mbox{G-type} stars. \textit{GALEX} observations show that the system is fairly bright in the ultraviolet. The ZTF light curve exhibits outbursts. We obtained a period of $10.79(4)$\,h from the \textit{TESS} light curve. This appears to be an U\,Gem dwarf nova, possibly with an evolved donor although an unevolved \mbox{G-type} donor would also be consistent.

\subsection{SDSS J080407.93--002216.4}

Gabriel Murawski first identified this as a CV from ZTF data (MGAB-V3549\footnote{https://sites.google.com/view/mgab-astronomy/full-list-mgab-v3501-v4000}). The SDSS spectrum shows single peaked asymmetric Balmer  and \ion{He}{ii} emission lines. The equivalent width of the  \ion{He}{ii} emission line is similar to that of the H$\beta$ line (Fig.\,\ref{fig:J0804-0022}).  The ZTF light curve shows orbital variations of $\Delta m\simeq2$ with a period of $1.56166(3)$\,h. Such a large amplitude can only be explained by cyclotron beaming, and this system is therefore a polar.

\subsection{SDSS J085210.44+783246.7}

The three SDSS spectra show  Balmer, \ion{He}{i} and  \ion{He}{ii} emission lines.   J0852+7832 is located within the novalike area of the HR diagram, and it is a moderately bright  \textit{GALEX} ultraviolet source. The ZTF and ASAS-SN light curves exhibit eclipses but no significant outbursts. We obtained a period of $17.1$\,h from the \textit{TESS} light curve and $17.109(1)$\,h from the ATLAS light curve.  We classify this system as a novalike albeit with an unusually long period.

\subsection{SDSS J090756.67+715859.2}

The SDSS spectrum shows double-peaked Balmer and  \ion{He}{i} emission lines above a blue continuum. The ASAS-SN and ZTF (Fig.\,\ref{fig:J0907+7158}) light curves show that the object is eclipsing with no outbursts. We found a period of $3.656$\,h from the \textit{TESS} light curve. Unpublished data from MDM together with the ZTF light curve reveal a period of $3.65597(2)$\,h. \citet{2022A&A...662A..40C} classified this object as a hot subdwarf but we consider that it is a novalike. The strong variability is unusual with the mass transfer rate varying around the critical value.  The light curve resembles that of the IW\,And subclass of novalikes. It is also similar to that of a Z\,Cam CV albeit that there are no ``standstills'' visible in the light curve.

\subsection{SDSS J092534.78+434917.3}

This system is also known as 2MASS J09253483+4349179. 
The SDSS spectrum shows single-peaked Balmer emission lines and also \ion{He}{ii} above a hot blue continuum. The ZTF and CRTS light curves do not show any outbursts.  A periodogram of the TESS light curve reveals three periods~--~a strong signal at  $3.434$\,h and weaker signals at $3.56(1)$\,h  and $4.04$\,d (Fig.\,\ref{fig:J0925+4349}). This would suggest that there is a warped disk giving rise to negative superhumps \citep{2015ApJ...803...55T}. Our periodogram seems very similar to that of V704\,And \citep{2022MNRAS.514.4718B} in which case  $3.56(1)$\,h would be the orbital period and $3.434$\,h the beat signal. J0925+4349 is located in the novalike area of the HR diagram, and based on the available evidence, we classify it  as a novalike.

\subsection{SDSS J160450.19+414328.3}

The SDSS spectrum shows single-peaked Balmer, \ion{He}{I} and \ion{He}{II} emission lines above a blue continuum. The object is located in the novalike area of the HR diagram. The ZTF light curve (Fig.\,\ref{fig:J1604+4143}) shows an extraordinary $\Delta m\simeq4$ dimming around $\mathrm{MJD}=59\,000$, lasting $\simeq\mathrm{six}$ months, indicative of J1604+4143 being of the VY\,Scl sub-class among the novalikes. Around $\mathrm{MJD}\simeq59800$, the system started to exhibit disc outbursts, resembling the behaviour of ASAS\,J071404+7004.3 (fig.\,2b in \citealt{2022MNRAS.510.3605I}). The steadily growing sample of known CVs and the availability of exquisite long-term photometric monitoring increasingly reveals the fluid boundaries between novalike and dwarf nova-like behaviour (e.g. \citealt{2019PASJ...71...20K}).

\subsection{SDSS J160811.26--135058.2}

This system is also known as USNO-A2.0 0750-09453854.
The SDSS spectrum shows double-peaked Balmer emission lines above the broad absorption lines from the white dwarf. No  signature of the donor is detected. The object sits on the white dwarf cooling sequence. Both the ZTF (Fig.\,\ref{fig:J1608-1350}) and CRTS light curves show frequent stunted  $\Delta m\simeq1$ outbursts. The spectrum and position on the HR diagram suggest that this object is a WZ\,Sge. However WZ\,Sge are characterised by rare, large, outbursts which is not consistent with the ZTF and CRTS data. We speculate that the stunted outbursts may be due to the object being weakly magnetic resulting in a truncated disc.

\subsection{SDSS J161817.28+030936.8}

This system is also known as BMAM-V636.
The SDSS spectrum shows Balmer, \ion{He}{i} and \ion{He}{ii} emission lines.  J1618+0309 is a known polar (BMAM-V636). Cyclotron humps are evident in the spectrum  (Fig.\,\ref{fig:J1618+0309}) indicating $B\simeq37$\,MG.

\subsection{SDSS J163805.39+083758.3}

This system is also known as V544 Her. The system is located just below  the main sequence in the HR diagram with a \mbox{K-type} donor visible in the SDSS spectrum. V544\,Her is also a UV-bright source detected by \textit{GALEX}, indicating significant flux from either an accretion disc or a hot white dwarf. Only weak H$\alpha$ emission is detected. The ZTF light curve shows at least six outbursts with $\Delta m\simeq5$. The periodogram of the ZTF light curve shows a number of aliases with the most likely period being $5.320(1)$\,h which we consider more probable than the tentative period of $1.66$\,h noted by \citet{1990PASP..102..758H}.

\subsection{SDSS J171247.71+604602.9}

The SDSS spectrum exhibits double-peaked Balmer and \ion{He}{i} emission lines and the SED shows that it is dominated by the \mbox{M-dwarf}. The object is located near to the main sequence in the HR diagram. The ZTF light curve shows  two  small outbursts and variability.   We find a period of $3.9167(1)$\,h from the ZTF light curve which is twice the period found by \citet{2020ApJS..249...18C}.  We believe the periodic behaviour is due to ellipsoidal modulation which occurs twice per orbit. We also have an MDM photometric time series corroborating the ZTF ellipsoidal period. The spectral type is  either an early M or late K, suggesting an evolved donor. An evolved donor would also be consistent with the absence of dwarf nova outbursts, as these systems seem to have lower accretion rates \citep{2022ApJ...934..142S}.

\subsection{SDSS J173905.58--452714.5}

This is the classical nova V728\,Sco, which erupted in 1862. The SDSS spectrum shows double-peaked Balmer, \ion{He}{i} and \ion{He}{ii} emission lines. J1739--4527 is located close to the main sequence in the HR diagram. There is only ATLAS light curve data available. We concur with \citet{2012MNRAS.423.2476T} that J1739--4527 now exhibits the characteristics of a ``normal'' CV, i.e. there is no spectroscopic signature related to the nova eruption left.

\subsection{SDSS J182319.26+051857.1}

This system is also known as Gaia17cbl.
The two SDSS spectra show double-peaked Balmer emission lines. Whereas the spectrum taken on $\mathrm{MJD}=59709$ was obtained during quiescence, the one taken one spectrum ($\mathrm{MJD}=59737$) shows a continuum flux level higher by a factor of about eight, and also shows a strong double-peaked \ion{He}{ii} emission line.  The ZTF and \textit{Gaia} light curves show frequent outbursts ($\Delta m\simeq2$). This is a U\,Gem, and the second SDSS spectrum was obtained during one of the outbursts.

\subsection{SDSS J191705.21--095144.8}

The SDSS spectrum shows multiple-peaked \ion{He}{i} emission lines indicating that this is an AM\,CVn. It is located on the white dwarf cooling sequence in the HR diagram. The ATLAS light curve appears to show four outbursts although subject to large uncertainties.

\subsection{SDSS J193436.19+510742.1}

This system is also known as V2289\,Cyg.
The SDSS spectrum shows double-peaked Balmer,  \ion{He}{i}   and \ion{He}{ii} emission lines. The object is located near the main sequence in the HR diagram. The ZTF light curve shows frequent ($P_\mathrm{outburst}\simeq20$\,d) outbursts. The \textit{TESS} light curve yields a period of  $3.8383(6)$\,h. We have a reliable spectroscopic orbital period of $3.41(5)$\,h from MDM, which implies that the photometric \textit{TESS} period arises from a superhump period. Using the relation between period excess and the mass ratio $q$ from \citet{2022arXiv220102945K} we calculate $q=0.46$ assuming that these are Stage A superhumps.  A closer examination of the ZTF light curve shows that most of the outbursts are actually superoutbursts with superhumps and so we classify this as an ER\,UMa, i.e. a dwarf nova with an extremely short super-outburst cycle. This is an unusually long period for an ER\,UMa, second only to SDSS\,J08084617+3131060 which has a period of $4.94$\,h.

\subsection{SDSS J193511.46--531745.8}

This system is also known as CRTS J193511.4-531746.
The SDSS spectrum shows single-peaked Balmer and \ion{He}{i} emission lines. The object is located within the main sequence in the HR diagram which is consistent with the SED which reveals a significant contribution of the donor star. The CRTS light curve shows at least seven outbursts. We obtained a tentative  period of $7.96(5)$\,h from the \textit{TESS} light curve consistent with the classification of U\,Gem.

\subsection{SDSS J200504.93+322122.5}

This is the SU\,UMa dwarf nova V550\,Cyg. The SDSS spectrum shows a red continuum with narrow Balmer absorption lines. The spectrum is consistent with the photometric SED and the synthetic magnitude is consistent with the \textit{Gaia} photometry. The object is located within the main sequence of the HR diagram when reddening of ($E(B-V)=0.69$) is accounted for. This is all inconsistent with  the properties of an SU\,UMa dwarf nova. CXOGSG\,J200505.2+322121has a textit{Chandra} X-ray detection in the range  $0.1-10$\,keV at 3.8\,arcsec away from the catalogue position of V550\,Cyg, and we tentatively identify \textit{Gaia}\,EDR3\,2031052139821246336 as the counterpart to this SU\,UMa dwarf nova, with a parallax of 2.83\,mas and a quiescent magnitude of $G\simeq20.8$. It is likely  that V550\,Cyg is actually \textit{Gaia}\,EDR3\,2031052139821246336 ~--~\citet{1999IBVS.4675....1S} refers to it being the ``southeastern of a pair'' and it is in a crowded area close to the Galactic plane.

\subsection{SDSS J200729.55+625047.6}

This system is also known as ZTF J200729.55+625047.6.
The SDSS spectrum shows  Balmer and \ion{He}{i} emission lines indicative of an accreting magnetic white dwarf. The spectrum and SED are dominated by the donor star. The object is located near the main sequence. The ZTF light curve (Fig.\,\ref{fig:J2007+6250}) shows frequent state changes but no apparent outbursts, which are consistent with our tentative classification of J2007+6250 as a magnetic CV. We obtained a period of $5.610679(1)$\,h from the \textit{TESS} light curve which we believe is more reliable than the $72$\,d period quoted by \citet{2020ApJS..249...18C}.

\section{Discussion}\label{sec:discussion}

\subsection{Effectiveness of the different CV targeting strategies within SDSS-V}
\begin{figure*} 
\centering
\includegraphics[width=\textwidth]{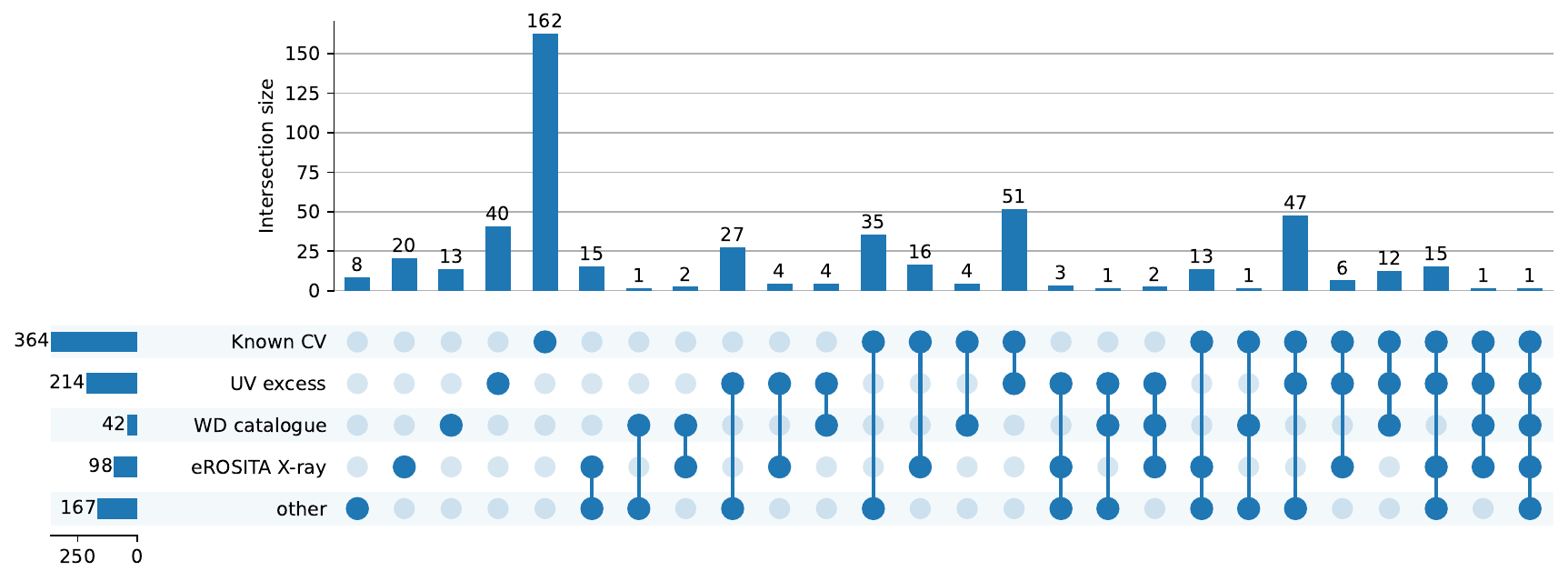}

\caption{\label{fig:UpSet} 
Each of the \NCVs CVs was targeted by one or more cartons. In this UpSet diagram \citep{6876017} the cartons have been grouped into five categories in order to identify any correlations. Known CVs include both published CVs and candidate CVs identified e.g. by their outbursts. Ultraviolet-excess objects have been targeted using data from  \textit{GALEX}, \textit{XMM} and \textit{Swift}.  The white dwarf catalogue category is focused on single white dwarfs and is essentially based on \citet{2021MNRAS.508.3877G}. The \textit{eROSITA} X-ray category includes candidate AGNs as well as compact binaries identified by their X-ray emission. The other category consists of targets from the spare fibre program.    }
\end{figure*}

In Section\,\ref{section:testing} the use of SDSS-V cartons was described noting that cartons can overlap in terms of their target contents. Figure\,\ref{fig:UpSet} illustrates the relative contribution of the targeting methods as implemented in the SDSS cartons that resulted in the identification of the \NCVs\ CVs. The left-most five histograms show the number of CVs that were identified only thanks to a \textit{single} carton, i.e. they were not part of any overlap between the different cartons. The other histograms to the right show the number of CVs that were targeted by multiple cartons, e.g. the right-most histogram illustrates that only a single CV was targeted by all cartons. It is interesting to note that, as expected, targeting known or candidate CVs, \texttt{mwm\_cb\_cvcandidates}, naturally accounted for the largest number (364) of CVs observed by \SDSSV, a large fraction of those ($\simeq44$~per cent) were not contained within any other carton. Inspection of these systems shows that this was because of the lack of ultraviolet photometry and/or poor \textit{Gaia} astrometry. The various ultraviolet-excess target selections (\texttt{mwm\_cb\_gaiagalex} and \texttt{mwm\_cb\_uvex1-5}) were also very successful in leading to the identification of CVs. Whereas the carton \texttt{mwm\_wd\_core} principally targets single white dwarfs selected from the \textit{Gaia} data \citep{2019MNRAS.482.4570G, 2021MNRAS.508.3877G}, it contains a number of low-accretion rate CVs with little to no spectroscopic signature of the donor star, primarily WZ\,Sge dwarf novae and AM\,CVn systems. Twenty CVs were observed by \SDSSV\ solely because of their properties in the \textit{eROSITA} survey with all but two arising from the \texttt{BHM\_SPIDERS\_AGN} cartons. These twenty were predominantly faint systems ($m>19$) which explains why eight did not have \textit{Gaia} counterparts and fourteen could not be completely classified. Our \textit{eROSITA}-related observations were all obtained based on an early release of \textit{eROSITA} data (eFEDS) with limited coverage and the proportion of \textit{eROSITA}-related CV discoveries is expected to rise following the larger eRASS1 release.  Finally, only eight CVs are entirely serendipitous discoveries. All eight were targeted by ``open fibre cartons'', different projects making use of spare fibres left over once all main survey targets are allocated in a given field. The specific open fibre programs in this case were a representative spectroscopic mapping of the \textit{Gaia} HR diagram and searches for quasars as well as for detached, spatially unresolved white dwarf plus main sequence star binaries. 

The very small fraction, $\simeq1.5$~per cent, that were not contained in any of the CV-specific plus \textit{eROSITA} cartons suggests that \SDSSV\ will deliver a very complete CV sample.

\subsection{Space density}

With two exceptions we have followed the approach described in Section\,6.4 of \citet{2023MNRAS.524.4867I} to estimate the space density of the CV sub-types. \citet{2023MNRAS.524.4867I} firstly calculated the limiting distance for each sub-type, within which \allsdss\ could reliably detect CVs of that sub-type. We use their limiting distances here. \citet{2023MNRAS.524.4867I} also assumed scale heights for CVs of each sub-type and we use the same assumptions here.   \citet{2023MNRAS.524.4867I} then estimated completeness from the proportion of a known sample of CVs that were rediscovered by \allsdss\ (we use a different approach here - see below). \citet{2023MNRAS.524.4867I} then determined the coverage of \allsdss\ taking account of galactic latitude by analysing the HEALPixs which were encompassed by one or more plates. We use the same approach here using the robotic fibre positioning system fields in place of plates. For each sub-type this results in the effective volume spanned by the survey footprint which is then used to estimate the space density.

There are two significant differences between \allsdss\ and \SDSSV\ which cause complications. The first is that \SDSSV\ deliberately targeted substantial areas within the Galactic plane whilst \allsdss\ focused on higher Galactic latitudes. The second complication is that \SDSSV\ deliberately targeted known CVs and objects with an ultraviolet or X-ray excess  whereas CVs from \allsdss\ were mostly serendipitous discoveries whilst searching for other types of objects. We consider the implications of these selection effects in the following two sub-sections.

\subsubsection{Targeting the Galactic plane}

A common assumption, also used by \citet{2023MNRAS.524.4867I}, is that the space density of CVs varies exponentially with height above or below the Galactic plane with the scale height dependent upon the age of the CVs. Estimating the space density of CVs in magnitude-limited surveys such as SDSS therefore has to take account of their height above the Galactic plane, which itself is a function of the typical age of the CVs in a given sub-class. 

\begin{figure} 
\centering
\includegraphics[width=\columnwidth]{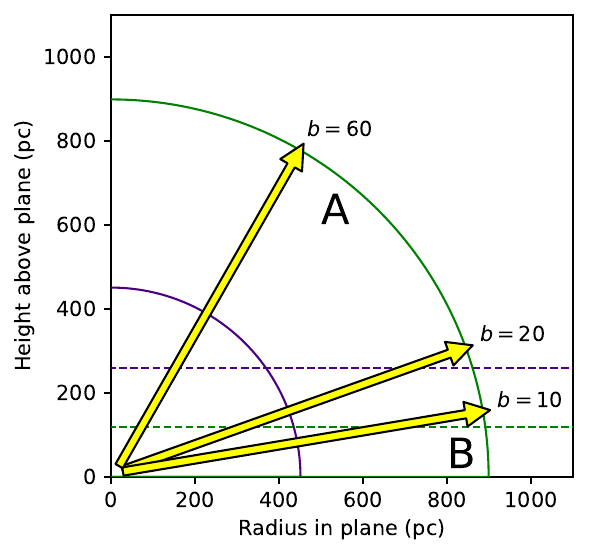}

\caption{\label{fig:galactic_lat}
Schematic portraying a vertical section through the Galactic disk with the Sun at the origin. The arcs depict the limiting distances (from \citealt{2023MNRAS.524.4867I}) up to which an SU\,UMa (indigo) and an U\,Gem (green) can be reliably detected by \SDSSV. The horizontal lines depict scale heights of $120$\,pc and $260$\,pc pertaining to U\,Gem and SU\,UMa respectively. The yellow arrows show the lines of sight for Galactic latitudes $|b|=10\degree,20\degree\, \mathrm{and}\,60\degree$.  Considering first a line of sight where $|b|=60\degree$ we see that there are very few U\,Gem CVs in region A as it is more than two scale heights above the plane.  Illustrating this with a quantitative example using typical values from Tables\,3\, and\,4 from \citet{2023MNRAS.524.4867I} we would predict number densities per steradian of viewing solid angle  of $N_\mathrm{U\,Gem}=\Veffugemsixty$\,$\mathrm{sr^{-1}}$\, and $N_\mathrm{SU\,UMa}=\Veffsuumasixty$\,$\mathrm{sr^{-1}}$. Considering now the case where $|b|=10\degree$ the SU\,UMas will be drawn from a similar range of distances as the first case  whilst proportionately far more U\,Gems will be observed from region B which is close to the plane and hence highly populated.  The quantitative example then predicts $N_\mathrm{U\,Gem}=\Veffugemten$\,$\mathrm{sr^{-1}}$\, and $N_\mathrm{SU\,UMa}=\Veffsuumaten$\,$\mathrm{sr^{-1}}$. This demonstrates the bias in favour of long-period CVs such as U\,Gem from observing lines of sight close to the Galactic plane.
}
\end{figure}

\begin{figure} 
\centering
\includegraphics[width=\columnwidth]{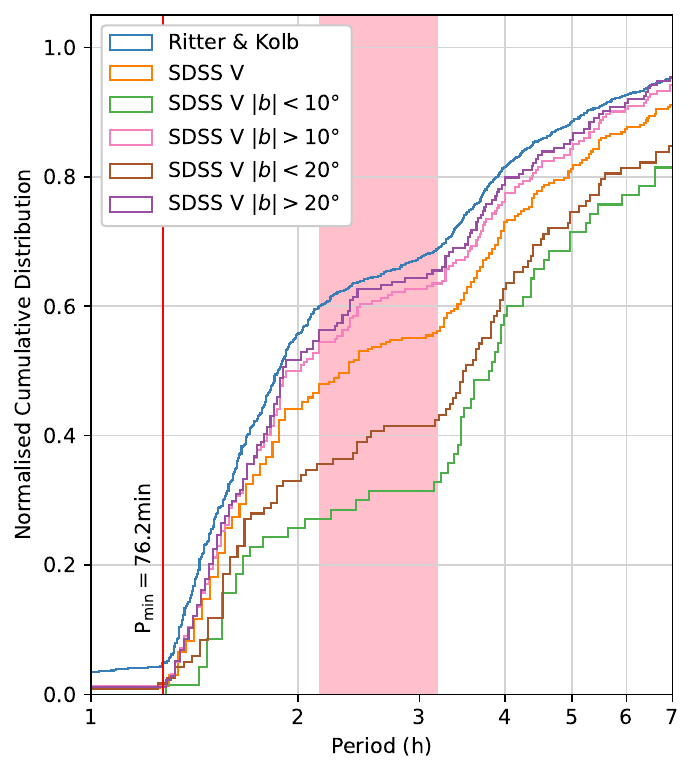}

\caption{\label{fig:periodcumdist}
Cumulative distribution of the periods of the CVs in \SDSSV. The distribution is compared with that of the Ritter and Kolb catalogue (version 7.24 \citealt{2003A&A...404..301R}). Subsets of the CVs in \SDSSV\, filtered by Galactic latitude are also shown which reveal the strong bias in favour of long-period CVs at low Galactic latitudes. The red shaded area is the ``period gap'' \citep{1980MNRAS.190..801W,2001ApJ...550..897H,2006MNRAS.373..484K}}
\end{figure}

Furthermore the limiting distance varies between sub-types according to their intrinsic absolute magnitude (which can vary by up to \textit{ten magnitudes})  implying that surveys at low Galactic latitudes are heavily skewed towards the detection of intrinsically bright systems thereby drastically changing the relative proportions of the different CV sub-types (Fig.\,\ref{fig:galactic_lat}). In addition, low Galactic latitudes are subject to distance-dependent reddening, which alters the limiting distance of a magnitude-limited survey as a function of the sight line. Further complications are caused by crowding in the Galactic plane resulting in less reliable \textit{Gaia} data (the key input into most of the CV targeting strategies) and the lack of coverage of the Galactic plane by \textit{GALEX}, greatly reducing the effectiveness of the ultraviolet-excess cartons. The overall effect of including observations near the Galactic plane is shown in Fig.\,\ref{fig:periodcumdist} where the bias in favour of long-period CVs is very evident at low Galactic latitudes. We address these issues by restricting the sample that we use to estimate the space density to those CVs with $|b|>20\degree$.

\subsubsection{Specifically targeting CVs}
In \allsdss, CVs were not specifically targeted and discoveries were essentially random. \citet{2023MNRAS.524.4867I} therefore assessed, and corrected for, the incompleteness of the CVs in \allsdss\ by considering the proportion of known CVs from \cite{2003A&A...404..301R} that were rediscovered in \allsdss. In \SDSSV\ we take a different approach. Figure\,\ref{fig:UpSet} shows that of the \SDSSV\ \NCVs CVs all but the eight ``other'' CVs were contained in one or more of the CV-related cartons from which we calculate that $98.4$\,per \,cent of the CVs in the footprint are targeted. Completeness is therefore determined by aggregating the  targets in the CV-related cartons and calculating the proportion that have actually been observed.

\subsubsection{Space density estimates}

\begin{table}
\centering
\caption{The completeness of a category of carton is the ratio of the number of objects observed to the number of targets within the coverage of \SDSSV. }
\label{tab:completeness}
\begin{tabular}{llll}
\hline
  Category & \begin{tabular}[c]{@{}l@{}}Objects \\ observed\end{tabular} & \begin{tabular}[c]{@{}l@{}}Targets in \\ Coverage\end{tabular} & Completeness \\ \hline
Known CV & 263 & 533 & 0.49 \\
UV excess & 49375 & 106359 & 0.46 \\
WD catalogue & 12321 & 24551 & 0.50 \\
\textit{eROSITA} x-ray & 67354 & 95268 & 0.71 \\
[0.5ex]
Total & 129313 & 226711 & 0.57
                        \\ \hline
\end{tabular}
\end{table}

\begin{figure*} 
\includegraphics[width=\textwidth]{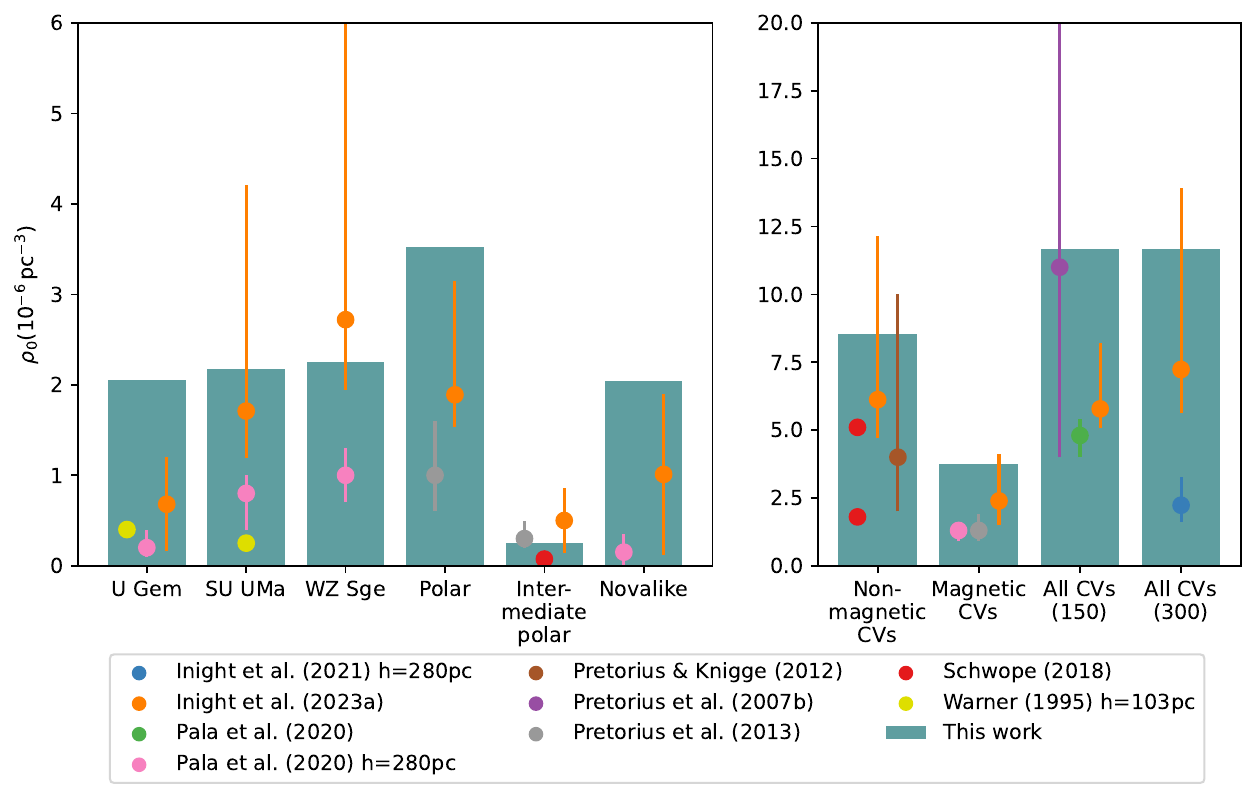}
\caption{\label{fig:space_densities} Comparison of the space density estimates for CVs from this work (bar chart) and published values. All estimates use the P2007 assumptions for scale height $h$ unless otherwise stated. Left panel: Estimates for individual sub-types. Note that the values from \citet{2020MNRAS.494.3799P} are understated as they do not take account of completeness.  Right panel: Composite estimates. \citet{2021MNRAS.504.2420I} specifically excluded selection effects.
 }   
\end{figure*}

We followed the approach in Section\,6.4 in \citet{2023MNRAS.524.4867I} using the subset of CVs from this work with $|b|>20\degree$ to avoid the Galactic plane. A footprint was established consisting of the sky coverage (defined in terms of HEALPixs) of all the fields with $|b|>20\degree$ containing at least one observation in a CV-related carton. This footprint took account of the different diameters in the focal plane of plates ($1.49\degree$) and fields at Apache Point observatory ($1.414\degree$) and Los Campanos observatory ($0.9465\degree$). Each carton was then analysed to find the total number of targets within the footprint and the proportion that had been observed and hence the overall completeness (see Table\,\ref{tab:completeness}). The assumptions from \citet{2023MNRAS.524.4867I} regarding scale heights and brightness were used to compute effective volumes and hence space densities. The uncertainty in the estimate for space density is dominated by the assumptions regarding scale height and we follow the approach used in \citet{2023MNRAS.524.4867I} to provide a notional error estimate using different scale height assumptions.  The results are tabulated in Table\,\ref{tab:spacedensity} and plotted in Fig.\,\ref{fig:space_densities}.  They are comparable with, and reinforce, previous findings \citep{1995cvs..book.....W,2007MNRAS.382.1279P,2012MNRAS.419.1442P,2013MNRAS.432..570P,2018A&A...619A..62S,2020MNRAS.494.3799P}.

\begin{table*}
\centering
\caption{Estimates of the space densities of the CV sub-types in \SDSSV\ for different assumptions of the scale height. $\rho_0$ (P2007) assumes the scale heights from \citet{2007MNRAS.374.1495P}. $N$ is the number of \SDSSV\, objects that are closer than the limiting distance. Note that the values of $\rho_0$ for individual sub-types are slightly understated ($\simeq20$\,per cent)  due to the remaining unclassified CVs and dwarf novae. 
This does not apply to the ``All CVs'' values which have been calculated for 150\,pc and 300\,pc limiting distances to enable comparison with \citet{2020MNRAS.494.3799P} and \citet{2021MNRAS.504.2420I}, respectively. These are calculated values and their precision should not be used to infer their uncertainty which is dominated by the uncertainty in the scale height (see text). }
\label{tab:spacedensity}
\begin{tabular}{lllllll} 
\hline
\multirow{2}{*}{Type} & \multirow{2}{*}{\begin{tabular}[l]{@{}l@{}}Limiting distance\\ (pc)\end{tabular}} & \multirow{2}{*}{$N$} & \multicolumn{4}{c}{Space Density $\mathrm{(10^{-6} pc^{-3})}$}       \\ 
\cline{4-7}
                      &                                                                                     &                    & $\rho_0(120)$ & $\rho_0(260)$ & $\rho_0(450)$ & $\rho_0(P2007)$            \\
\hline
    U Gem & 899 & 29 & 2.07 & 0.54 & 0.30 & 2.07 \\
SU UMa & 451 & 28 & 4.70 & 2.09 & 1.52 & 2.19 \\
WZ Sge & 412 & 20 & 4.38 & 2.16 & 1.63 & 2.27 \\
Polar & 256 & 10 & 5.37 & 3.33 & 2.78 & 3.54 \\
Intermediate polar & 739 & 3 & 0.26 & 0.08 & 0.05 & 0.26 \\
Novalike & 2676 & 42 & 2.05 & 0.24 & 0.07 & 2.05 \\
[0.5ex]
All CVs & 150 & 6 & 13.89 & 10.62 & 9.61 & 11.71 \\
All CVs & 300 & 44 & 15.41 & 8.69 & 6.97 & 11.71  
    \\\hline
\end{tabular}
\end{table*}

\section{Conclusions}\label{sec:conclusions}

We have used a convolutional neural network (CNN) to analyse the complete set of spectroscopic data from \SDSSV\,resulting in the discovery of \Nnew new CVs, spectroscopic confirmation of \Nspec CV candidates and \Nnewperiods\ new or improved orbital periods. We have demonstrated that the CNN recovers at least 90~per cent of CVs within a sample whilst rejecting at least 98~per cent of the non-CVs. Furthermore we have shown that, with the exception of pre-polars, there is no significant selection bias between CV sub-types. This approach can be refined, using our enlarged sample of known CVS for training, and used not only for future \SDSSV\ surveys but potentially 
 also for other spectroscopic surveys such as DESI, WEAVE and 4MOST. 

We have also used our set of CVs to estimate the space density of each sub-type building upon previous estimates.  

We have analysed the effectiveness of the CV-related cartons used in \SDSSV\ and shown that they are very effective at targeting CVs  as long as the required input data (parallaxes, UV and X-ray photometry) is available. This situation is expected to improve following the full implementation of \textit{eROSITA} and future missions such as \textit{UVEX}.

\section{Acknowledgements}
Funding for the Sloan Digital Sky Survey V has been provided by the Alfred P. Sloan Foundation, the Heising-Simons Foundation, and the Participating Institutions. SDSS acknowledges support and resources from the Center for High-Performance Computing at the University of Utah. The SDSS web site is \url{www.sdss5.org}. SDSS is managed by the Astrophysical Research Consortium for the Participating Institutions of the SDSS Collaboration, including the Carnegie Institution for Science, Chilean National Time Allocation Committee (CNTAC) ratified researchers, the Gotham Participation Group, Harvard University, The Johns Hopkins University, L'Ecole polytechnique f\'{e}d\'{e}rale de Lausanne (EPFL), Leibniz-Institut f\"{u}r Astrophysik Potsdam (AIP), Max-Planck-Institut f\"{u}r Astronomie (MPIA Heidelberg), Max-Planck-Institut f\"{u}r Extraterrestrische Physik (MPE), Nanjing University, National Astronomical Observatories of China (NAOC), New Mexico State University, The Ohio State University, Pennsylvania State University, Smithsonian Astrophysical Observatory, Space Telescope Science Institute (STScI), the Stellar Astrophysics Participation Group, Universidad Nacional Aut\'{o}noma de M\'{e}xico, University of Arizona, University of Colorado Boulder, University of Illinois at Urbana-Champaign, University of Toronto, University of Utah, University of Virginia, Yale University, and Yunnan University. 

The authors are honored to be permitted to conduct astronomical research on Iolkam Du’ag (Kitt Peak), a mountain with particular significance to the Tohono O’odham.

Based on observations obtained with the Samuel Oschin 48$-$inch Telescope at the Palomar Observatory as part of the Zwicky Transient Facility project. ZTF is supported by the National Science Foundation under Grant No. AST-1440341 and a collaboration including Caltech, IPAC, the Weizmann Institute for Science, the Oskar Klein Center at Stockholm University, the University of Maryland, the University of Washington, Deutsches Elektronen-Synchrotron and Humboldt University, Los Alamos National Laboratories, the TANGO Consortium of Taiwan, the University of Wisconsin at Milwaukee, and Lawrence Berkeley National Laboratories. Operations are conducted by COO, IPAC, and UW

CRTS is supported by the U.S. National Science Foundation under grants AST-0909182 and CNS-0540369. The work at Caltech was supported in part by the NASA Fermi grant 08-FERMI08-0025, and by the Ajax Foundation. The CSS survey is funded by the National Aeronautics and Space Administration under Grant No. NNG05GF22G issued through the Science Mission Directorate NEOs Observations Program. 

This research has made use of the International Variable Star Index (VSX) data base, operated at AAVSO, Cambridge, Massachusetts, USA. 

This paper includes data collected by the TESS mission. Funding for the TESS mission is provided by the NASA's Science Mission Directorate.

This work has made use of data from the European Space Agency (ESA) mission {\it Gaia} (\url{https://www.cosmos.esa.int/gaia}), processed by the {\it Gaia} Data Processing and Analysis Consortium (DPAC, \url{https://www.cosmos.esa.int/web/gaia/dpac/consortium}). Funding for the DPAC has been provided by national institutions, in particular the institutions participating in the {\it Gaia} Multilateral Agreement.  This research has made use of NASA's Astrophysics Data System. This research has made use of the VizieR catalogue access tool, CDS, Strasbourg, France.

This work has made use of data from the Asteroid Terrestrial-impact Last Alert System (ATLAS) project. The Asteroid Terrestrial-impact Last Alert System (ATLAS) project is primarily funded to search for near earth asteroids through NASA grants NN12AR55G, 80NSSC18K0284, and 80NSSC18K1575; byproducts of the NEO search include images and catalogs from the survey area. This work was partially funded by Kepler/K2 grant J1944/80NSSC19K0112 and HST GO-15889, and STFC grants ST/T000198/1 and ST/S006109/1. The ATLAS science products have been made possible through the contributions of the University of Hawaii Institute for Astronomy, the Queen’s University Belfast, the Space Telescope Science Institute, the South African Astronomical Observatory, and The Millennium Institute of Astrophysics (MAS), Chile.

This project has received funding from the European Research Council (ERC) under the European Union’s Horizon 2020 research and innovation programme (Grant agreement No. 101020057).  MRS acknowledges support from Fondecyt (grant 1221059). G.T. was supported by the IN109723  grant from the Programa de Apoyo a Proyectos de Investigación e Innovación Tecnológica (PAPIIT).

\section{Data Availability}
\SDSSV\, data will be publicly available at the end of the proprietary period. 
The other data used in this article are available from the sources referenced in the text.


\nocite{*}

\bibliographystyle{mnras}

\bibliography{refs,others,refs_supp}



\newpage

\appendix 
\section{Convolutional Neural network (CNN) model}\label{sec:appendix1}

This section is intended to provide enough information to enable the reader to reproduce our work and also to explain the rationale for the design. This inevitably takes a high level and simplified approach to many aspects of machine learning that are beyond the scope of this paper. 

Our objective was to achieve a small shortlist of potential CVs whilst minimising the number of false positives. However to avoid introducing additional and unquantifiable biases we deliberately limited the training data to spectra. This effectively replicated the manual process whereby spectra are superficially scanned by eye to generate shortlists. Other data such as photometry and astrometry can be used in the subsequent analysis. We used the \textsc{python keras } package to implement the CNN. 

\subsection{Model assessment}

The design of a CNN consists of choosing a number of components together with values for the associated hyperparameters. A metric for the performance of a given configuration is therefore needed to guide these choices. We chose the following approach (see Chapter 3 in \citealt{géron2022hands} for further details). 
When a CNN is applied to the spectrum of a known CV it will either predict that it is a CV, hence a true positive (TP), or that it is not, hence a false negative (FN). Similarly when it is applied to the spectrum of a known non-CV it will either predict that it is a CV, hence a false positive (FP), or that it is not, hence a true negative (TN). In practice the CNN calculates a number which can be regarded as the probability that the spectrum is a CV. This leads to the use of a threshold to convert the probability to a binary decision. 
The threshold is a variable parameter and two related functions that depend upon it are the true positive rate (TPR) and the false positive rate (FPR):

\begin{equation}
TPR= \frac{N_{\mathrm{TP}}}{\left(N_{\mathrm{TP}}+N_{\mathrm{FN}}\right)}
\end{equation}
\begin{equation}
FPR= \frac{N_{\mathrm{FP}}}{\left(N_{\mathrm{FP}}+N_{\mathrm{TN}}\right)}
\end{equation}

\begin{figure} 
\centering
\includegraphics[width=\columnwidth]{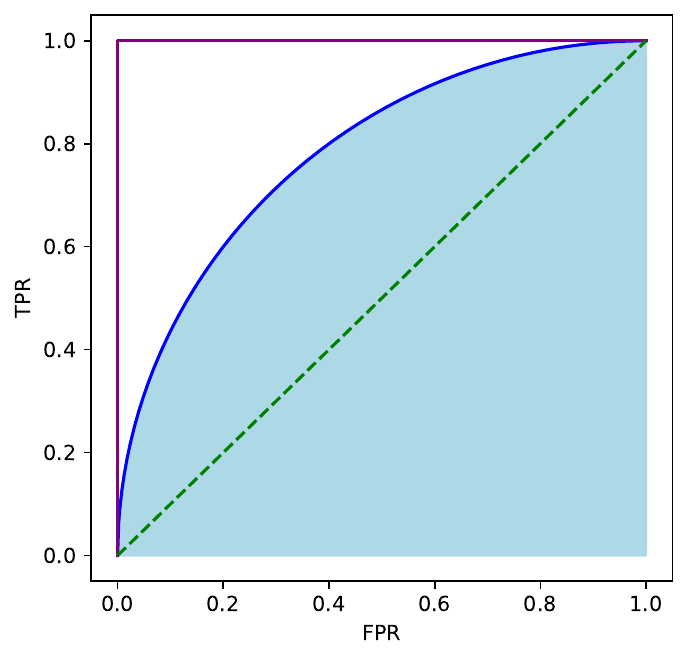}
\caption{\label{fig:ROC} An illustration of Receiver Operating Characteristic (ROC) curves showing the ideal curve (purple) and a typical curve (blue). The green dotted curve represents an untrained CNN that is effectively making random predictions. The Area Under the Curve (AUC) for the typical case is shaded light blue. 
    }
\end{figure}

If the threshold is close to zero both the TPR and the FPR will also be close to zero. Similarly if the threshold is close to one both the TPR and the FPR will be close to one (Fig.\,\ref{fig:ROC}). A plot of TPR against FPR is called the Receiver Operating Characteristic (ROC) curve ( Fig.\,\ref{fig:ROC}).  Ideally as the threshold is increased the TPR will increase more rapidly than the FPR with a perfect CNN exhibiting the purple curve in  Fig.\,\ref{fig:ROC}. Conversely a CNN delivering a random result would exhibit the green curve in  Fig.\,\ref{fig:ROC}.  It follows that the shaded Area Under the Curve (AUC) in   Fig.\,\ref{fig:ROC} is a measure of the quality of the CNN and we use AUC in the following discussion.

\subsection{Model structure}

\begin{figure} 
\centering
\includegraphics[width=0.6\columnwidth]{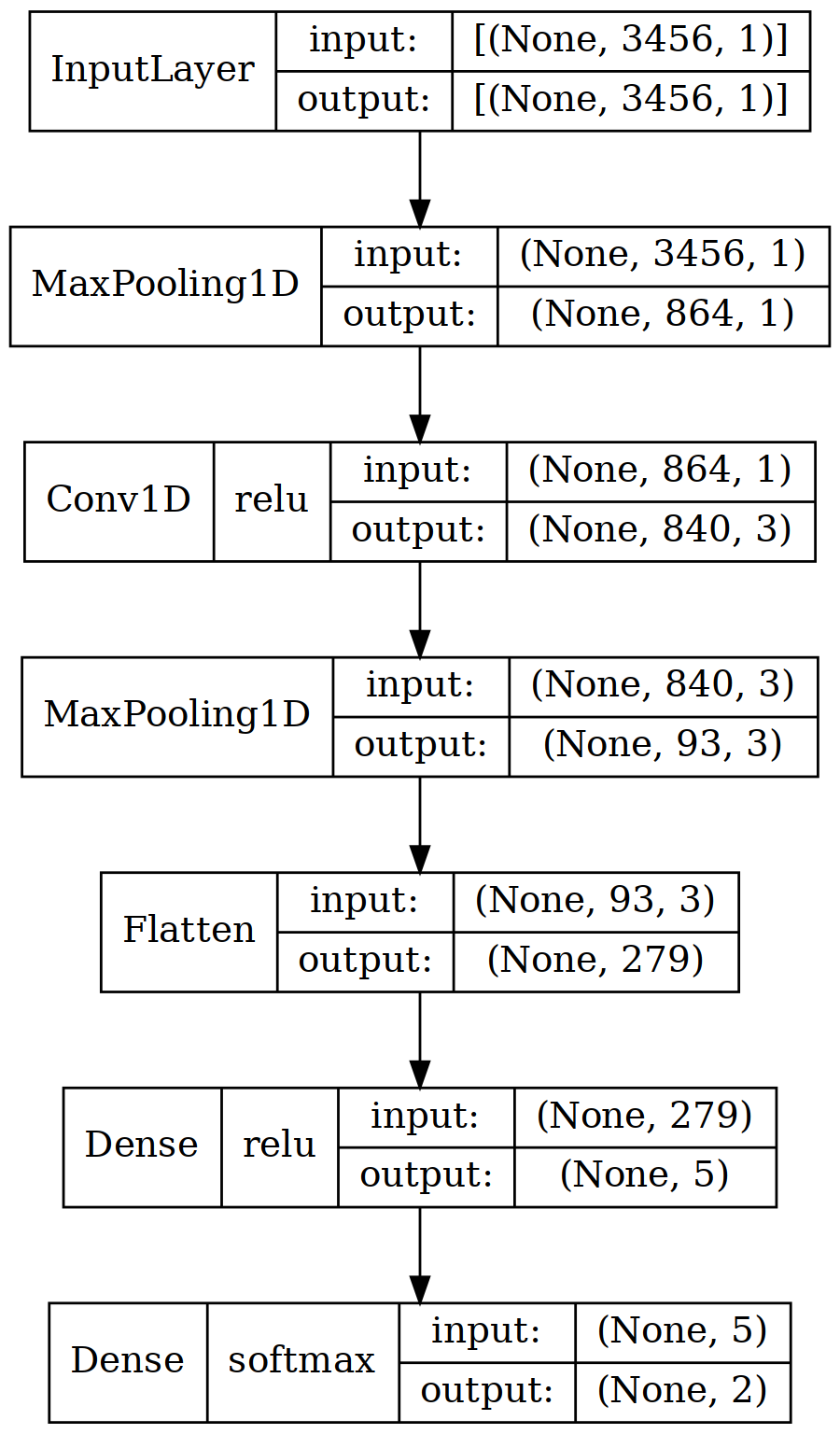}

\caption{\label{fig:final_CNN}
The final model. The input layer purely acts as a buffer. The pooling layer aggregates the outputs of the input layer with each output being the maximum of four inputs. Each output of the convolution layer is a function of a subset of the inputs. The layer has four kernels ($\mathrm{width}=25$) and uses the rectified linear unit function (ReLU) as its activation function.  The second pooling layer aggregates nine inputs for every output and the flattening layer then reshapes the data into one dimension. Each dense layer is fully mapped with every output the weighted average of every input. An activation function is then applied - ReLU in the first layer and softmax in the second.    }
\end{figure}

The final model together with a brief explanation is shown in Fig.\,\ref{fig:final_CNN}. A number of experiments were performed to evaluate the effect of changing the configuration and hyperparameters.

\subsubsection{The number of convolutional and dense layers}

The effect of increasing the number of convolutional and dense layers was investigated (Table\,\ref{Table:layers}). When allowance has been made for stochastic errors introduced by the modelling process there is no significant improvement to justify the increased complexity and computational effort. This is not unexpected as the number of inputs (i.e. the 3456 wavelength bins) is small by comparison with the number of free parameters (the trainable weights) in the model.    

 \begin{table}
 \caption{\label{Table:layers} The effect of adding a second convolutional or dense layer to the CNN. The AUC is virtually unchanged.  }
 \centering
 \begin{tabular}{ccc}
  \hline
Number of & Number of & AUC \\
convolutional&dense&\\
layers&layers&\\
\hline 
 1 &1& 0.9976 \\
1& 2& 0.9977 \\
2& 1& 0.9984 \\
2& 2& 0.9983 \\
 \hline
 \end{tabular}
 \end{table}

\subsubsection{The number of kernels, and kernel width, in the convolution layer.}

\begin{figure} 
\centering
\includegraphics[width=\columnwidth]{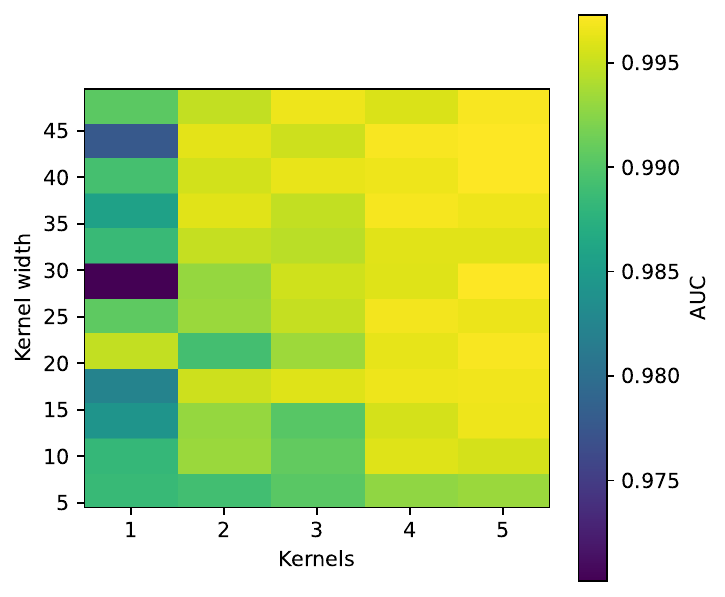}
\caption{\label{fig:kernels and widths} Heat map showing the effect of increasing the number of kernels and the width of each kernel. The stochastic variation in the output of the CNN model is very evident even though each estimate was repeated five times and the median taken. The choice of three kernels and a width of 25 results is a reasonable compromise between quality and overfitting.
   }
\end{figure}

The number of kernels and the choice of kernel width largely determine the number of free parameters in the model and can potentially result in overfitting.  Fig.\,\ref{fig:kernels and widths} shows how increasing the number of kernels to three improves the AUC whilst the kernel width appears to have little influence once it exceeds 15.

\subsubsection{The number of nodes in the dense layers}

\begin{figure} 
\centering
\includegraphics[width=\columnwidth]{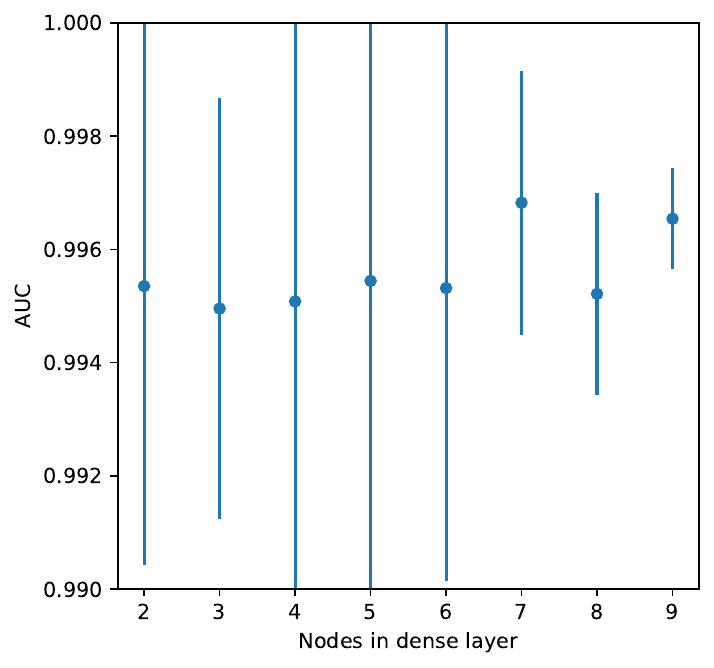}

\caption{\label{fig:dense_nodes}
The effect on AUC of the number of nodes in the first dense layer. There is considerable stochastic variation in the output of the CNN for given input data and so we repeated the analysis ten times and took the median value of AUC. The error bars are based upon the standard deviation of the results of the individual analyses.    }
\end{figure}

The number of nodes in the first dense layer is a compromise between quality and computational effort. It was found (Fig.\,\ref{fig:dense_nodes}) that increasing the number of nodes beyond five did not result in a material improvement in AUC.

The number of nodes in the second dense layer has to be two as this is a binary classifier.

\subsubsection{The proportion of non-CV spectra in the test/train dataset}\label{sec:pnoncv}

\begin{figure} 
\centering
\includegraphics[width=\columnwidth]{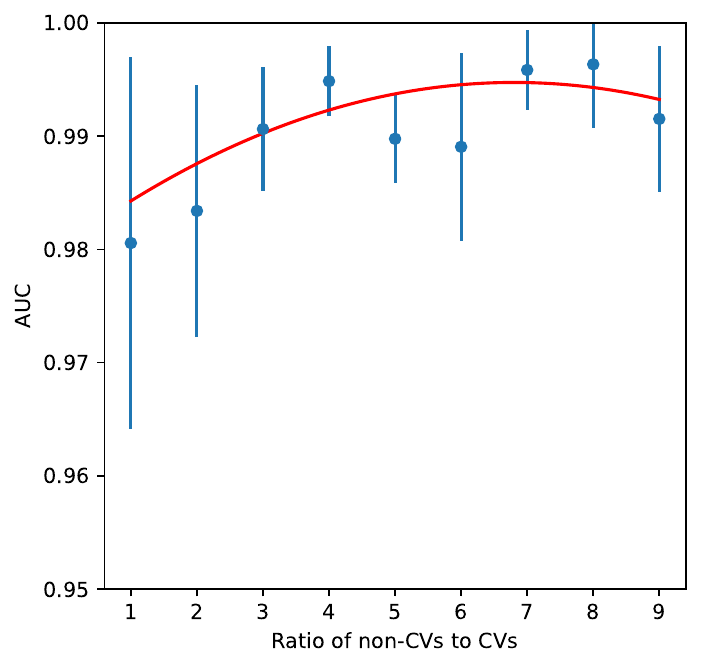}

\caption{\label{fig:proportion_of_non_CVs}
The effect on AUC of the ratio of CVs to non-CVs in the train/test dataset. The datapoints were produced by progressively increasing the ratio of non-CV data to CV data in the train/test dataset in each case running the CNN 25 times and then taking the median AUC. It is apparent that adding further labelled non-CVs to the train/test dataset will not improve the quality of the CNN.   }
\end{figure}

The size of the test/train dataset is limited by the number of known labelled CVs. Whilst in principle the spectra of a very large number of non-CVs could be added to the test/train dataset there will be a point at which this is not worthwhile. We experimented with the ratio of CVs to non-CVs by progressively adding additional non-CVs to the test/train dataset and creating a new model. The results are shown in Fig.\,\ref{fig:proportion_of_non_CVs} and demonstrate that there is no value in increasing the ratio beyond $\simeq10$ and that our train/test dataset is optimal. 

\subsection{Training}

\begin{figure} 
\centering
\includegraphics[width=\columnwidth]{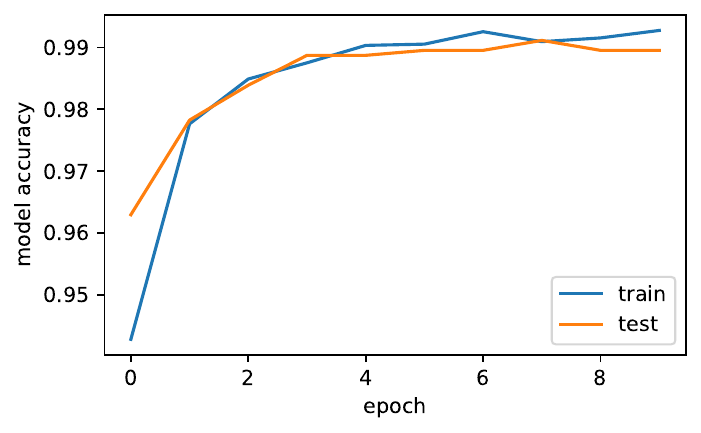}
\includegraphics[width=\columnwidth]{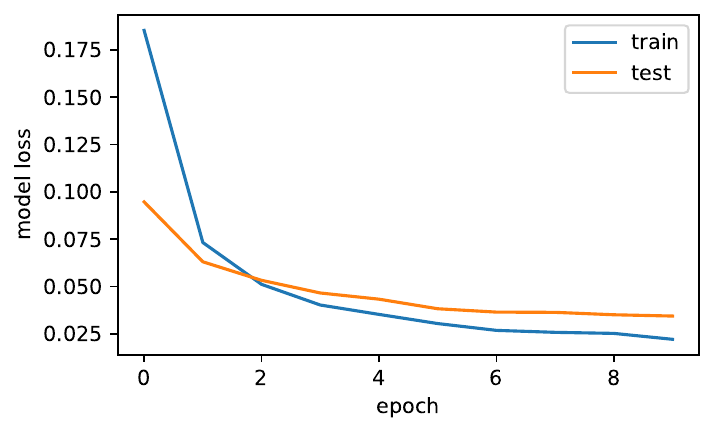}
\caption{\label{fig:History of accuracy and loss} 
The accuracy (top panel) and loss (bottom panel) of the training and test data following each epoch of iteration. This demonstrates that continuing the iteration beyond ten epochs will not improve the accuracy and only cause overfitting.  }
\end{figure}

The CNN was trained on the train/test dataset by iteratively improving the solution where each iteration is called an epoch. After ten epochs the accuracy had reached  $99.2$\,per cent. Similarly the loss of the test data levelled off whilst the loss of the training data continued to fall (see Fig.\,\ref{fig:History of accuracy and loss}). Iteration was stopped after ten epochs to prevent over-training.

 \subsection{Evaluation of the CNN}

We split the train/test dataset into separate samples with 20 per cent being reserved for test data. The remaining 80 per cent was then used to train the CNN. 

\begin{figure} 
\centering
\includegraphics[width=\columnwidth]{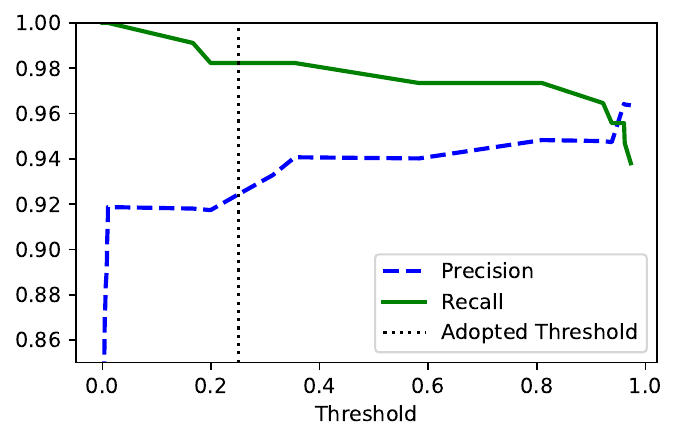}
\caption{\label{fig:bimodal probability} 
Precision and recall as a function of threshold. Neither precision nor recall are very sensitive to the choice of threshold, as the distribution of the probabilities of being a CV is highly bimodal. We chose a threshold of $0.25$  as a reasonable compromise between missing some CVs (false negatives leading to lower recall) and an excessive number of candidates (false positives leading to lower precision).    }
\end{figure}

When trained, the final CNN acts as a predictor and will calculate the probability of a given spectrum being a CV. A threshold is then used to make a binary classification of being a CV or not. The choice of threshold affects the proportion of CVs recovered. In practice, this choice is not very sensitive as the distribution of probability is highly bimodal (see Fig.\,\ref{fig:bimodal probability}). 

We have therefore used a threshold of $0.25$ as being a reasonable compromise between the number of false positives (non-CVs wrongly predicted to be CVs) and the number of false negatives (CVs wrongly predicted to be non-CVs).


\bsp	
\label{lastpage}
\end{document}